\documentclass[journal]{IEEEtran}
\usepackage{multirow}
\usepackage{booktabs}
\usepackage{amsfonts}
\usepackage{amsmath}
\usepackage{amsmath,cases}
\usepackage{amssymb}
\usepackage{graphicx}
\usepackage{cite}
\usepackage{epsfig}
\usepackage{url}
\usepackage{algorithm}
\usepackage{algorithmicx, algpseudocode}
\usepackage{epstopdf}
\usepackage{color}
\usepackage{mathrsfs}
\usepackage[justification=centering]{caption}
\usepackage{float}
\usepackage{diagbox}
\usepackage{subcaption}
\usepackage{balance}

\newcommand{\clA}{{\cal A}}

\newcommand{\la}{\langle}
\newcommand{\ra}{\rangle}

\newcommand{\bX}{\boldsymbol{X}}

\newcommand{\clL}{{\cal L}}
\newcommand{\clN}{{\cal N}}

\newcommand{\clK}{{\cal K}}

\newcommand{\diag}{{\sf diag}}

\newcommand{\clB}{{\cal B}}

\newcommand{\bz}{\mathbf{z}}

\newcommand{\clH}{{\cal H}}

\newcommand{\bY}{\mathbf{Y}}

\newcommand{\zi}{z^{(\iota)}}

\newcommand{\zio}{z^{(\iota+1)}}

\newcommand{\Xiota}{X^{(\iota)}}

\newcommand{\bV}{\mathbf{V}}

\newcommand{\riota}{r^{(\iota)}}
\newcommand{\ai}{a^{(\iota)}}

\newcommand{\Vi}{V^{(\iota)}}
\newcommand{\Vio}{V^{(\iota+1)}}

\newcommand{\bi}{b^{(\iota)}}

\newcommand{\tri}{\tilde{r}^{(\iota)}}

\newcommand{\Psii}{\Psi^{(\iota)}}
\newcommand{\chii}{\chi^{(\iota)}}

\newcommand{\btheta}{\pmb{\theta}}

\newcommand{\thetai}{\theta^{(\iota)}}

\newcommand{\tvarphii}{\tilde{\varphi}^{(\iota)}}

\newcommand{\thetaio}{\theta^{(\iota+1)}}

\newcommand{\tfi}{\tilde{f}^{(\iota)}}

\newcommand{\clHi}{\clH^{(\iota)}}
\newcommand{\tbz}{\tilde{\bz}}

\newcommand{\Yi}{Y^{(\iota)}}
\newcommand{\Upsiloni}{\Upsilon^{(\iota)}}
\newcommand{\clBi}{\mathcal{B}^{(\iota)}}
\newcommand{\clCi}{\mathcal{C}^{(\iota)}}
\newcommand{\tclCi}{\tilde{\mathcal{C}}^{(\iota)}}
\newcommand{\Omegai}{\Omega^{(\iota)}}

\newcommand{\gi}{g^{(\iota)}}

\newcommand{\Xiii}{\Xi^{(\iota)}}
\newcommand{\sigmai}{\sigma^{(\iota)}}
\newcommand{\tiH}{\tilde{H}}
\newcommand{\tbtheta}{\tilde{\btheta}}
\newcommand{\tF}{\tilde{F}}
\newcommand{\clO}{\mathcal{O}}
\allowdisplaybreaks
\begin{document}
\title{A New Class of Analog Precoding for Multi-Antenna Multi-User Communications over High-Frequency Bands}
\author{W. Zhu$^{1}$, H. D. Tuan$^1$, E. Dutkiewicz$^1$,  H. V. Poor$^2$, and L. Hanzo$^3$
	\thanks{The work was supported in part by the Australian Research Council's Discovery Projects under Grant DP190102501,  in part by the U.S National Science Foundation under Grants CNS-2128448 and ECCS-2335876, in part by  the Engineering and Physical Sciences Research Council projects EP/W016605/1, EP/X01228X/1 and EP/Y026721/1 as well as of the European Research Council's Advanced Fellow Grant QuantCom (Grant No. 789028)}
	\thanks{$^1$School of Electrical and Data Engineering, University of Technology Sydney, Broadway, NSW 2007, Australia (email: wenbo.zhu@student.uts.edu.au, tuan.hoang@uts.edu.au, eryk.dutkiewicz@uts.edu.au); $^2$Department of Electrical and Computer Engineering, Princeton University, Princeton, NJ 08544, USA (email: poor@princeton.edu);
		$^3$School of Electronics and Computer Science, University of Southampton, Southampton, SO17 1BJ, U.K (email: lh@ecs.soton.ac.uk) }
}
\date{}
\maketitle
\begin{abstract}
A network relying on a large antenna-array-aided base station is designed for delivering  multiple information streams to multi-antenna users over high-frequency bands such as the millimeter-wave and sub-Terahertz bands.
The state-of-the-art analog precoder (AP) dissipates  excessive circuit power  due to its reliance on a large number of phase shifters.
To mitigate the power consumption,
we propose a novel  AP relying on a controlled number of phase shifters. Within this new AP framework, we  design a hybrid precoder (HP) for maximizing the users'  minimum throughput, which poses a computationally challenging problem of large-scale, nonsmooth mixed discrete-continuous log-determinant optimization. To tackle this challenge, we develop an algorithm which iterates through solving
convex problems  to generate a sequence of HPs that converges to the max-min solution. We also introduce a new framework of smooth optimization termed soft max-min throughput optimization. Additionally,
we develop another algorithm, which iterates by evaluating closed-form expressions to generate a sequence of HPs that converges to the soft max-min solution. Simulation results reveal that the HP soft max-min solution approaches the Pareto-optimal solution {\color{black} constructed for simultaneously optimizing both the minimum throughput and sum-throughput. Explicitly, it achieves a minimum throughput similar to directly maximizing the users' minimum throughput and it also attains a sum-throughput similar to directly maximizing the sum-throughput.}
\end{abstract}
\begin{IEEEkeywords}
Millimeter-wave and terahertz bands, multi-stream delivery, power-efficiency, hybrid precoding, analog precoding, digital precoding,  log-determinant optimization, mixed discrete continuous optimization.
\end{IEEEkeywords}

\section{Introduction}
Wireless communications over high-frequency bands, including the millimeter-wave (mmWave) and the sub-Terahertz (sTHz) \cite{Rapetal19}, as well as the Terahertz (THz) bands \cite{SN11,Akyetal22,TD23}, are currently considered the only viable means of meeting the demands of high-volume data delivery in next generation communication networks and beyond.
To compensate for the substantial path loss associated with these frequency bands, it is necessary to harness  a large number of transmit antennas at the base station (BS) for signal transmission.
Hybrid precoding (HP) composed of analog precoding (AP) and baseband digital precoding (DP) plays a pivotal role as the essential signal processing technique designed for focusing the desired signal and for mitigating the interference at the receiver end.  It is worth noting that AP relies on radio frequency chains (RFs) of phase shifters, which can result in excessive circuit   power consumption. As an illustrative example, an AP with $8$ RFs connected to $144$ transmit antennas requires $144\times 8=1152$ phase shifters in a full-connected (FC) structure, where each of the $8$ RFs is connected to all transmit antennas. Alternatively,  $144$ phase shifters may be harnessed in an array-of-subarray (AoSA) structure \cite{Gaoetal16}, where each of the $8$ RFs
is connected to $144/8=16$ transmit antennas only. For a circuit power consumption of $20$ mW  per phase shifter \cite{Winetal13}, such an AP consumes $1152\times 20=23040$ mW under FC, and $144\times 20=2880$ mW under AoSA,  making its implementation infeasible. Additionally, it is important to note that the circuit power  consumption of a phase shifter increases significantly in higher frequency bands. Given the current state-of-the-art in HP, which requires a number of phase shifters that is no less than that of transmit antennas, it appears unlikely that extra-high bandwidth communications relying on extra-large antenna arrays can be practically implemented in the foreseeable future.

Another notable challenge in the realm of wireless communications over high-frequency bands is that they are mostly studied within the context of either a single multi-antenna user \cite{Ayetal14,Yuetal16,LLW17,YHY20,DTCP22,SDN23,Chenetal23} or multiple single-antenna users \cite{SY16,KHY18,Shi-18-Jun-A,Nasetal20TVT,FMLS21,SAA21,Yuetal23tvt,Yuetal23twc}. This {\color{black}previous} focus tends to overlook the more practical scenario of multiple multi-antenna users.
In single-user communication, the interference is negligible, making high-volume data delivery relatively straightforward. When dealing with multiple single-antenna users, improving their throughputs is equivalent to improving their signal-to-interference-plus-noise ratios (SINRs), which can be effectively addressed using fractional programming. However, when it comes to enhancing the throughputs of multi-antenna users, things become considerably more intricate. The user throughputs are the log-determinant (log-det) functions of nonlinear matrix expressions composed of signal and interference covariances. This intricate enhancement presents a unique matrix optimization challenge that cannot be adequately addressed using the conventional tools of applied optimization.

\begin{table*}[!htb]
	\centering
	\caption{ Highlighting the distinctive contributions in comparison to the related literature}
	\begin{tabular}{|l|c|c|c|c|c|c|c|c|}
		\hline
		\backslashbox{Contents}{Literature} & \textbf{This work}&\cite{Ayetal14,Yuetal16,LLW17,YHY20,DTCP22,SDN23,Chenetal23} &\cite{KHY18,FMLS21}&\cite{SY16,KHY18,Shi-18-Jun-A}
		& \cite{Nasetal20TVT} & \cite{Yuetal23tvt,Yuetal23twc}\\
		\hline
		Single multi-antenna user & &$\surd$& & & &\\
		\hline
		Multiple single-antenna users & &   & $\surd$& $\surd$ & $\surd$ &$\surd$\\
		\hline
		Cubic complexity& &   &        &    & $\surd$ &\\
		\hline
		Scalable complexity&$\surd$ &   & &  & &$\surd$\\
		\hline
		Fair quality of delivery &$\surd$ &   &  & & $\surd$ &\\
		\hline
		Multiple multi-antenna users &$\surd$ & & &  &  &\\
		\hline
		Power consumption efficiency &$\surd$ &  & &  & & \\
		\hline
		Phase shifter control &$\surd$ &  & &  &  &\\
		\hline
	\end{tabular}
	\label{tab:CompaIRSon}
\end{table*}

The aim of the present paper is to introduce a new class of AP to control the number of phase shifters for maintaining a realistic power consumption for wireless communications over high-frequency bands. Our contributions are as follows:
\begin{itemize}
\item We develop a new class of APs, which control the number of phase shifters by linking any phase shifter to multiple antennas. As a result, each RF component can efficiently exploit a controlled number of phase shifters as their link to a predefined set of antennas;
\item Within this new class of  APs, our primary focus is on designing  a HP for maximizing the minimum throughput of multi-antenna users, thereby ensuring uniform quality of information delivery (QoD) to all users. Again, the user throughputs  are characterized by
    the log-det function of nonlinear matrix expressions  involving both signal and interference covariances. Due to the practical constraints of low-resolution phase shifters, maximizing this minimum log-det function poses a significant computational challenge within the context of nonsmooth, large-scale mixed discrete optimization. To tackle this challenge,  we develop a nonsmooth max-min log-det algorithm that iterates by solving convex problems to generate a sequence of HPs, which  converges towards the max-min solution. Extensive numerical simulations  demonstrate that under the same transmit power budget, HP within this new AP framework achieves the users' minimum  throughput comparable to that delivered by  HP associated with the conventional AP, which consumes much more circuit power by utilizing excessive numbers of phase shifters for implementation. Remarkably, under the same power consumption,  the HPs associated  with this new AP attain a much higher users' minimum  throughput than that achieved by HP using the conventional AP. In fact, the minimum user throughput attained by the latter is notably lower compared
     to that achieved by the former;
\item The computational complexity  associated with  solving convex problems in the nonsmooth max-min log-det algorithm is a cubic function of the number of decision variables, resulting in particularly high computational demands due to the large scale of these convex problems.  To mitigate this issue, we introduce the concept of  soft max-min log-det optimization. The concept aims for maximizing a soft and smooth approximation of the nonsmooth  minimum  log-det function. We develop a soft max-min log-det algorithms  that iterates by evaluating closed-form
    expressions having scalable complexity to generate a sequence of HPs, which converges toward the soft max-min solution. Notably, numerical simulations demonstrate that this HP solution is Pareto-optimal in the context of multi-objective optimization. Explicitly, it not only achieves a users'  minimum   throughput similar to that attained by direct max-min throughput optimization, but also has a sum-throughput similar to that achieved by direct sum-throughput maximization. Consequently, soft max-min throughput optimization presents a new HP design framework for computationally efficient multi-objective optimization.
\end{itemize}

Table  \ref{tab:CompaIRSon} demonstrates the advancements and distinctive contributions of this work in comparison to the existing related literature.

The  paper is organized as follows. Section II introduces a new class of APs designed for mitigated power consumption.  Section III focuses on the computational solution of designing HP to maximize the users' minimum throughput, while Section IV addresses the computational solution of designing HPs for  maximizing the  soft users'  minimum throughput.  Section V provides numerical simulation results, while the appendix provides some mathematical ingredients for the developments in  Sections III and IV.

\emph{Notation.\;} Only the optimization variables are boldfaced. For a complex number $x$,
 $\angle x$ presents its argument.
 Then $e^{\jmath x}\triangleq (e^{\jmath x_1}, \dots, e^{\jmath x_N})^T\in\mathbb{C}^N$
 for $x=(x_1,\dots, x_N)^T\in\mathbb{R}^N$. The inner product between vectors $x$ and $y$ is defined as $\langle x,y\rangle=x^Hy$.
 Analogously, $\la X,Y\ra\triangleq {\sf trace}(X^HY)$ for the matrices $X$ and $Y$.
 ${\sf diag}[X_1,\dots,X_N]$ is a block diagonal  matrix of the diagonal blocks
 $X_n$, $n=1,\dots, N$, while $\begin{bmatrix}X_{m,n}\end{bmatrix}_{(m,n)\in{\cal M}\times {\cal N}}$ or $\begin{bmatrix}X(m,n)\end{bmatrix}_{(m,n)\in{\cal M}\times {\cal N}}$ is a partitioned
 matrix of the submatrices
 $X_{m,n}$ or $X(m,n)$. We also use $\la X\ra$ for the trace of $X$ when $X$ is a square matrix. $X\succeq 0$ ($X\succ 0$, resp.) means that $X$ is Hermitian symmetric and positive semi-definite (positive definite, resp.). Accordingly, $X\succeq Y$ ($X\succ Y$, resp.) means that
$X-Y\succeq 0$ ($X-Y\succ 0$, resp.). $||X||$ is the Frobenius norm of the matrix $X$, which is defined by $\sqrt{\la X^HX\ra}$. $[X]^2$ stands for $XX^H\succeq 0$, so $||X||^2=\la [X]^2\ra$. Then $\ln X$ is the natural logarithm of the determinant (log-det) of $X\succ 0$. Whenever $X\succeq 0$,
$\sqrt{X}$ is a positive semi-definite matrix such that $[\sqrt{X}]^2=X$ and $\sqrt{X}(m,n)$ is
the $(m,n)$-th entry of $\sqrt{X}$.
$1_N\in \mathbb{R}^N$ is a $N$-dimensional vector with all entries equal to $1$, while $I_N$ is the $N\times N$ identity matrix. When the size of the identity matrix is  clear from the context, we
may omit the subscript $N$ in expressions.
Lastly, $\mathbb{R}^{N}_+$ is the set of
$N$-dimensional real vectors with positive entries.

\section{New analog precoders based on reduced numbers of phase shifters}
Let us consider a downlink (DL) scenario, where  a base station (BS) serves $K$ users, each
identified   by $k\in \clK\triangleq \{1, 2,\dots, K\}$. In this set up, the BS is equipped with a massive $N_e\times N_a$-element circular cylindrical array, while each user (UE) $k$ is equipped with an $N_t$-antenna array. Thus, the total number of transmit antennas is $N\triangleq N_eN_a$.

\begin{figure}[!t]
	\centering
	\begin{subfigure}[c]{0.4\textwidth}
		\centering
		\includegraphics[width=0.75\textwidth]{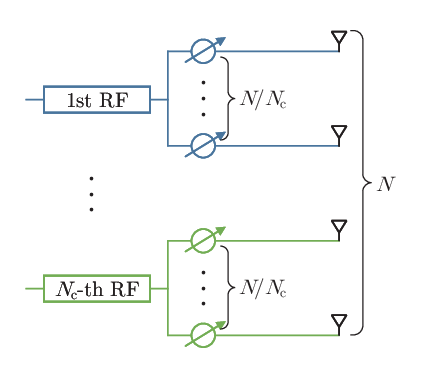}
		\caption{Conventional AoSA structure}
		\label{fig:AoSA_structure}
	\end{subfigure}
	\hfill
	\begin{subfigure}[c]{0.4\textwidth}
		\centering
		\includegraphics[width=1\textwidth]{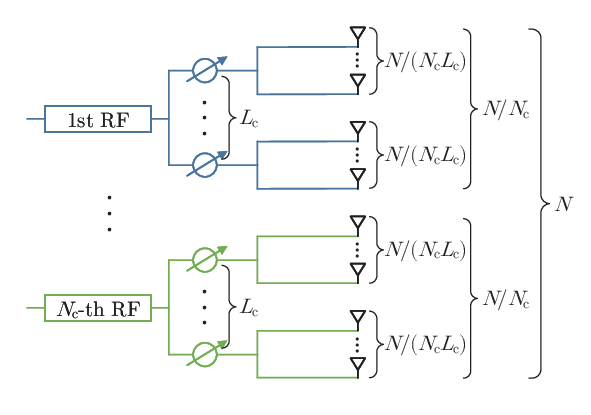}
		\caption{nAoSA structure}
		\label{fig:new_AoSA_structure}
	\end{subfigure}
	\caption{$a)$ The conventional AoSA: each of the $N/N_c$ phase shifter of an RF is connected to a single antenna; $b)$ The new AoSA (nAoSA): each of the $L_c$ phase shifters of an RF is connected to a set
of $N/(L_cN_c)$ antennas.}
	\label{fig:AoSA_nAoSA_structures}
\end{figure}

Let $N_c$ be the number of RF chains that the  BS  uses for its AP. Considering the BS as an $N$-antenna array, we partition it into $N_c$-antenna subarrays, with each of which  comprising $L=N/N_{c}$ antennas, forming what is known as an array of subarrays (AoSA) \cite{AHRP13}. Subsequently,
the $n_c$-th RF chain is connected to the $n_c$-th subarray, where  $n_c\in\clN_c\triangleq \{1,
\dots, N_c\}$ (see Fig. \ref{fig:AoSA_structure}). The
AP is defined by a matrix $\bV_{A}\in \mathbb{C}^{N\times N_c}$ in the following form
\begin{equation}\label{sa1}
\bV_A\triangleq {\sf diag}[\tbz_{n_c}]_{n_c\in\clN_c},
\end{equation}
with
\begin{equation}\label{bphi}
\tbz_{n_c}\triangleq  \left(\tbz_{n_c,1}, \dots, \tbz_{n_c,L}\right)^T=e^{\jmath\tbtheta_{n_c}}\in\mathbb{C}^L, n_c\in \clN_c,
\end{equation}
for
\begin{equation}\label{br1c}
\tbtheta_{n_c}\triangleq \left(\tbtheta_{n_c,1}, \dots, \tbtheta_{n_c,L}\right)^T\in \mathbb{R}^L.
\end{equation}
The AoSA-AP  $\bV_A$ (\ref{sa1}) relies on $N$ phase shifters formulated as $e^{\jmath \btheta_{l,n_c}}$,
$l\in\clL\triangleq \{1,\dots, L\}; n_c\in\clN_c$ for implementation. More precisely, the $l$-th phase shifter $e^{\jmath \btheta_{l,n_c}}$
in the $n_c$-th RF chain is linked to the $l$-th antenna within the $n_c$-th subarray.
In essence, each RF chain relies on $L$ phase shifters for its connection to an $L$-antenna subarray.
Recall that, with $20$ mW being the power consumption per phase shifter \cite{Winetal13}, for $N=12\times 12=144$ and $N_c=8$, the  AoSA (\ref{sa1}) already consumes
$2880$ mW.

Our aim is to propose a new AP structure that relies on significantly fewer phase shifters to facilitate practical implementation.
To reduce the total number $N=LN_c$ of phase shifters for implementing the AoSA-AP $\bV_A$ in (\ref{sa1}), we have to assign
$\tbz_{n_c}\in\mathbb{C}^L$ in (\ref{bphi}) a specific structure, which enables the $n_c$-th RF chain to rely on  reduced number $L_c<<L$
of phase shifters for its connection to the $n_c$-th subarray.
 Let $\clA\in \mathbb{R}^{L\times L_c}$, so that for each $\ell\in\clL\triangleq
\{1,\dots, L\}$, we have $\clA(\ell,\ell_c)=1$ only for a single
$\ell_c\in\clL_c\triangleq \{1,\dots, L_c\}$ and $\clA(\ell,\ell'_c)=0$ for all other $\ell'_c\neq \ell_c$. {\color{black} It is noteworthy that the matrix $\clA$ offers flexibility in selecting an arbitrary value for $L_c$, as it effectively maps the reduced number $L_c$ of phase shifters to $L$ phase shifters in the conventional AoSA-AP structure. Therefore, it plays a crucial role in developing our new AP structure.}
We propose the following structure defined to as the new AoSA (nAOSA):
\begin{equation}\label{naosa1}
\tbz_{n_c}=\clA\bz_{n_c}, n_c\in\clN_c,
\end{equation}
with
\begin{equation}\label{naosa2}
\bz_{n_c}\triangleq  \left(\bz_{n_c,1},	\dots,	\bz_{n_c,L_c}\right)^T
=e^{\jmath\btheta_{n_c}}\in\mathbb{C}^{L_c}, n_c\in \clN_c,
\end{equation}
for
\begin{equation}\label{naosa3}
\btheta_{n_c}\triangleq \left(\btheta_{n_c,1}, \dots, \btheta_{n_c,L_c}\right)^T\in\mathbb{R}^{L_c},
\end{equation}
i.e. the nAoSA AP obeys
\begin{equation}\label{naosa1a}
\bV_A\triangleq {\sf diag}[\clA\bz_{n_c}]_{n_c\in\clN_c}.
\end{equation}
By letting
\begin{equation}\label{ztheta}
\bz\triangleq \begin{bmatrix}
		\bz_{1}\cr
		\dots\cr
		\bz_{N_c}
	\end{bmatrix}\in\mathbb{C}^{N_cL_c}, \btheta\triangleq \begin{bmatrix}
		\btheta_{1}\cr
		\dots\cr
		\btheta_{N_c}
	\end{bmatrix}\in\mathbb{R}^{N_cL_c},
\end{equation}
we can rewrite (\ref{naosa2}) as
\begin{equation}\label{naosa2e}
\bz=e^{\jmath \btheta}.
\end{equation}
This structure allow the $n_c$-th RF chain to rely only on  $L_c$ phase shifters for its connection to
the $n_c$-th subarray. Specifically, each phase shifter
$e^{\jmath \btheta_{n_c,\ell_c}}$ in the $n_c$-th RF chain
is connected to the $\ell$-th antenna of $n_c$-th subarray  if $\clA(\ell,\ell_c)=1$.
{\color{black} For example, consider a scenario associated with $L=8$ and $L_c=4$, where each RF chain uses $4$ phase shifters to connect to an $8$-antenna subarray. Given that each phase shifter is linked to two adjacent antennas, $\clA$ can be formulated as:}
\begin{equation}\label{cla1}
\clA=\diag[1_2, 1_2, 1_2, 1_2]\in\mathbb{R}^{8\times 4},
\end{equation}
{\color{black} where $1_2$ denotes a $2$-dimensional vector with all entries equal to $1$. Here,} the $\ell_c$-th phase shifter $e^{\jmath \btheta_{n_c,\ell_c}}$, $\ell_c\in \clL_c$ within the $n_c$-th RF chain
is linked to two antennas: the $(L(n_c-1)+2\ell_c-1)$-th and the
$(L(n_c-1)+2\ell_c)$-th antennas of the $n_c$-th subarray.
 On the other hand, {\color{black}considering that each phase shifter is linked to two spaced antennas, $\clA$ can be alternatively formulated as:}
\begin{equation}\label{cla2}
\clA=\begin{bmatrix}I_4\cr
I_4
\end{bmatrix}\in\mathbb{R}^{8\times 4},
\end{equation}
{\color{black} where $I_4$ denotes the $4\times 4$ identity matrix. Here,} the $\ell_c$-th phase shifter $e^{\jmath \btheta_{n_c,\ell_c}}$ within the $n_c$-th RF chain is linked to two antennas: the $(L(n_c-1)+\ell_c)$-th and the
$(L(n_c-1)+\ell_c+4)$-th antennas  of the $n_c$-th subarray. Through numerical simulations, we have found that the structure (\ref{cla1}) outperforms its
counterpart (\ref{cla2}).   Fig. \ref{fig:AoSA_nAoSA_structures} contrasts
the conventional AoSA structure and our proposed nAoSA structure{\color{black}, where Fig. \ref{fig:new_AoSA_structure} shows the nAoSA structure associated with $\clA$ defined by (\ref{cla1}).}

Given that ABF relying on phase-shifters having an infinite resolution  is impractical for mmWave communication \cite{KYW13}, we opt for phase-shifters with of low $b$-bit resolution, i.e.
\begin{equation}\label{br1}
\btheta_{n_{c},\ell_c}\in \clB\triangleq \{ b'\frac{2\pi}{2^b}, b'=0, 1, \dots, 2^b-1\}.
\end{equation}
In what follows,  the projection of $\alpha\in [0,2\pi)$ into $\clB$ denoted by $\lfloor\alpha\rceil_b$ is referred to as its $b$-bit rounded version:
\begin{equation}\label{br2}
\lfloor\alpha\rceil_b=\nu_{\alpha}\frac{2\pi}{2^b}
\end{equation}
with
\begin{equation}\label{br3}
\nu_{\alpha}\triangleq \mbox{arg}\min_{\nu=0, 1,\dots, {\color{black}2^b}}\left|\nu \frac{2\pi}{2^b}-\alpha\right|,
\end{equation}
which can be readily found, because we have $\nu_{\alpha} \in \{\nu, \nu+1\}$ for $\alpha\in [\nu \frac{2\pi}{2^b}, (\nu+1)\frac{2\pi}{2^b}]$. {\color{black}We reset $\nu_{\alpha}= 0$, when it is $2^b$.}
When $b=\infty$, it is true that $\alpha=\lfloor \alpha\rceil_{\infty}$.
\section{Hybrid precoding design under nAoSA-AP}
We encode $n\in\clN\triangleq \{1,\dots, N\}$
by $(n_e(n),n_a(n))\in\clN_e\times\clN_a$, so that $n=n_a(n)N_e+n_e(n)$, and let
\begin{align}\label{chanmod2}
\mathbb{C}^{N_t\times N}\ni H_{k}=\begin{bmatrix}H_{k,1}&\dots&H_{k,N_{c}}\end{bmatrix},\nonumber\\
H_{k,n_c}=\begin{bmatrix}H_{k,n_c}(1)\cr
\dots\cr
H_{k,n_c}(N_t)
\end{bmatrix}\in\mathbb{C}^{N_t\times L}
\end{align}
be the channel  between  the  BS  and UE $k\in\clK$.
Let $s_k\in C(0,I_{N_t})$ be the information stream  intended for UE $k$, which is precoded by the HP
$\bV_A\bV_k$,
where $\bV_A$ is the nAOSA-AP defined from (\ref{naosa1a}), and
\begin{equation}\label{Vb}
\mathbb{C}^{N_c\times N_t}\ni \bV_{k}\triangleq \begin{bmatrix}\bV_{k}(1,1)&\dots&\bV_{k}(1,N_t)\cr
\dots&\dots&\cr
\bV_{k}(N_c,1)&\dots&\bV_{k}(N_c,N_t)
\end{bmatrix}
\end{equation}
is the digital baseband precoder (DP). The transmit signal at the BS is formulated as
$\sum_{k'\in\clK}\bV_A\bV_{k'}s_{k'}$.
The signal received at UE $k$ is
\begin{equation}\label{hb3}
\mathbb{C}^{N_t}\ni y_k=H_k\sum_{k'\in\clK}\bV_A\bV_{k'}s_{k'}+n_k,
\end{equation}
where $n_k\in C(0,\sigma I_{N_t})$ with $\sigma>0$ represents the noise, which  includes
the background noise and channel  error due to imperfect channel estimation.

It follows from (\ref{hb3}) that  the signal received at UE $k$ under the nAoSA AP $\bV_A$ (\ref{naosa1a}) is
\begin{equation}
\mathbb{C}^{N_t}\ni y_k=\clH_k(\bz)\sum_{k'\in\clK}\bV_{k'}s_{k'}
+n_k, \label{pgs}
\end{equation}
where we have
\begin{align}\label{clhi}
\clH_{k}(\bz)&\triangleq H_{k}{\sf diag}[\clA\bz_{n_c}]_{n_c\in\clN_c}\nonumber\\
&=\begin{bmatrix}H_{k,1}{\color{black}\clA}\bz_{1}&\dots&H_{k,N_c}{\color{black}\clA}\bz_{N_c}\end{bmatrix}
\in\mathbb{C}^{N_t\times N_{c}}, k\in\clK.
\end{align}
We will also use the representations (\ref{sa7a})--(\ref{sa8}) shown at the top of next page.
\begin{figure*}[t]
\begin{align}
	\clH_{k}(\bz)\bV_{k'}&=\sum_{n_c\in\clN_c}\begin{bmatrix}H_{k,n_c}(1)\clA\bz_{n_c} \bV_{k'}(n_c,1) &\dots&
H_{k,n_c}(1)\clA\bz_{n_c} \bV_{k'}(n_c,N_t) \cr
\dots&\dots&\dots\cr
H_{k,n_c}(N_t)\clA\bz_{n_c} \bV_{k'}(n_c,1) &\dots&
H_{k,n_c}(N_t)\clA\bz_{n_c} \bV_{k'}(n_c,N_t)
\end{bmatrix}\nonumber\\
&=\sum_{n_c\in\clN_c}\begin{bmatrix}\bV_{k'}(n_c,1)H_{k,n_c}(1)\clA\bz_{n_c}  &\dots&
 \bV_{k'}(n_c,N_t) H_{k,n_c}(1)\clA\bz_{n_c}\cr
\dots&\dots&\dots\cr
\bV_{k'}(n_c,1)H_{k,n_c}(N_t)\clA\bz_{n_c}  &\dots&
 \bV_{k'}(n_c,N_t)H_{k,n_c}(N_t)\clA\bz_{n_c}
\end{bmatrix}\nonumber\\
&=\begin{bmatrix}\tiH_{k,\ell,\ell'}(\bV_{k'})\bz\end{bmatrix}_{(\ell,\ell')\in\clN_t\times\clN_t}\label{sa7a}\\
&\triangleq \tiH_{k}(\bV_{k'},\bz) \label{sa7}
\end{align}	
for
\begin{equation}\label{sa8}
 \tiH_{k,\ell,\ell'}(\bV_{k'})\triangleq
 \begin{bmatrix}		\bV_{k'}(1,\ell)H_{k,1}(\ell'){\color{black}\clA}&\dots&\bV_{k'}(N_c,\ell)H_{k,N_c}(\ell'){\color{black}\clA}
	\end{bmatrix}.
\end{equation}
\hrulefill
\vspace{-0.4cm}
\end{figure*}
{\color{black} For $\bV\triangleq \{\bV_k, k\in\clK$\},} the  throughput of UE $k$ is defined by the following log-det function:
\begin{align}
r_k(\bV,\bz)&\triangleq {\color{black}\ln\left|I_{N_t}+[\clH_k(\bz)\bV_k]^2
\Psi_k^{-1}(\bV,\bz)\right|}\label{chanmode4}\\
&={\color{black}\ln\left|I_{N_t}+[\tiH_k(\bV_k,\bz)]^2
\Psi_k^{-1}(\bV,\bz)\right|}\label{chanmode5}
\end{align}
with
\begin{align}
\Psi_k(\bV,\bz)&\triangleq\sum_{k'\neq k}[\clH_k(\bz)\bV_{k'}]^2+\sigma I_{N_t}\label{chanmode6}\\
&=\sum_{k'\neq k}[\tiH_k(\bV_{k'},\bz)]^2+\sigma I_{N_t}.\label{Psiie}
\end{align}
Given the transmit power budget $P$, the BS's transmit power is  constrained as
\begin{align}
	&\sum_{k\in\clK}||{\sf diag}[\clA\bz_{n_c}]_{n_c\in\clN_c}\bV_{k}||^2= L\sum_{k\in\clK}||\bV_{k}||^2\leq P\nonumber\\
	\Leftrightarrow&\sum_{k\in\clK}||\bV_{k}||^2\leq P_L,\label{sa6}
\end{align}
for $P_L\triangleq P/L$, which is  independent of $\bz$.

To ensure a uniform quality-of-delivery (QoD) for all users in terms of their throughputs, our HP design is based on  the following problem of max-min log-det function  optimization:
\begin{equation}\label{basic}
\max_{\bz,\btheta,\bV} f(\bV,\bz)\triangleq \min_{k\in\clK}r_k(\bV,\bz) \quad\mbox{s.t.}\quad  (\ref{naosa3}), (\ref{naosa2e}), (\ref{br1}), (\ref{sa6}).
\end{equation}
The objective function in (\ref{basic})  is notably intricate and nonsmooth due to its reliance on the point-wise minimum of the nonlinear log-det functions $r_k(\bV,\bz)$, $k\in\clK$ involving the matrix variables $\bV_k$, $k\in\clK$ and vector variable $\bz$. Furthermore, the  constraints (\ref{naosa3}), (\ref{naosa2e}), (\ref{br1}) in (\ref{basic})   exhibit a blend of highly nonlinear and mixed discrete-continuous characteristics.  More particularly, the presence of the nonlinear equality constraint (\ref{naosa2e}) makes alternating optimization in  $\bz$ and $\btheta$ with
the other variables held fixed still computationally intractable. Consequently, (\ref{basic})  poses a formidable computational challenge within the domain of nonsmooth optimization.
To address this challenge, we embrace  the popular penalized optimization framework of
\cite{Betal06,PTKD12,CTN14,Tametal16,Yeetal20,Yuetal20jsac} to have the following penalized optimization reformulation for (\ref{basic}):
\begin{align}\label{sa11}
	\max_{\btheta,\bV,\bz} F_{\gamma}(\bV,\bz,\btheta)\triangleq \left[ f(\bV,\bz) -\gamma||\bz-e^{\jmath \btheta}||^2\right]\nonumber\\ \mbox{s.t.}\quad (\ref{br1}), (\ref{sa6}),
\end{align}
where $\gamma>0$ is a penalty factor {\color{black} introduced to integrate the nonlinear equality constraint (\ref{naosa2e}) into the optimization objective function}. Note that the problem (\ref{sa11}) is free from the nonlinear
equality constraint (\ref{naosa2e}), {\color{black} and a feasible point for (29) might not be automatically feasible for (28), unless the penalty term in the objective function in (29) is zero.}
As we will see shortly, in contrast to (\ref{basic}), its penalized optimization reformulation (\ref{sa11}) facilitates computationally tractable alternating optimization in either $\bz$ or $\btheta$, with
the other variables held fixed, although $\btheta$ is a discrete variable.

Initialized by
$(V^{(0)}, z^{(0)}, \theta^{(0)})$ feasible for (\ref{sa11}), let $(\Vi,\zi,\thetai)$ be a feasible point for (\ref{sa11}) that is found from the $(\iota-1)$-st iteration. The alternating optimization procedure at the $\iota$-th iteration to generate $(\Vio,\zio,\thetaio)$  feasible for (\ref{sa11})
so that
\begin{equation}\label{fial}
F_{\gamma}(\Vio,\zio,\thetaio)> F_{\gamma}(\Vi,\zi,\thetai)
\end{equation}
unfolds as follows.
\subsection{DP alternating optimization}
{\color{black} In the $\iota$-th DP alternating optimization round, we seek the next feasible point $\Vio$, while keeping $(\btheta,\bz)$ fixed at $(\thetai,\zi)$.} To generate $\Vio$ so that
\begin{align}\label{fial1}
&F_{\gamma}(\Vio,\zi,\thetai)> F_{\gamma}(\Vi,\zi,\thetai)\nonumber\\
\Leftrightarrow &f(\Vio,\zi)>f(\Vi,\zi),
\end{align}
we consider the following problem of  alternating optimization in DP $\bV$ for (\ref{sa11}) with $(\btheta,\bz)$ held fixed at $(\thetai,\zi)$:\footnote{As the penalty term
 and the constraint (\ref{br1}) in (\ref{sa11}) are independent on $\bV$, they are omitted during the alternating optimization in $\bV$}
\begin{equation}\label{bo}
	\max_{\bV}f(\bV,\zi)\triangleq \min_{k\in\clK}\riota_{1,k}(\bV)\quad \mbox{s.t.}\quad (\ref{sa6}),
\end{equation}
where according to (\ref{chanmode4}) and (\ref{chanmode6}):
\begin{align}\label{bmgm2}
\riota_{1,k}(\bV)&\triangleq r_k(\bV,\zi)\nonumber\\
&={\color{black}\ln\left|I_{N_t}+[\clHi_{1,k}\bV_{k}]^2(\Psii_{1,k}(\bV))^{-1}\right|}
\end{align}
for
\begin{equation}\label{bo4}
\Psii_{1,k}(\bV)\triangleq \Psi_k(\zi, \bV)=
\sum_{k'\neq k}\la \clHi_{1,k},[\bV_{k'}]^2\ra+\sigma I_{N_t},
\end{equation}
with
\begin{equation}\label{bo5}
\clHi_{1,k}\triangleq \clH_k(\zi), k\in\clK.
\end{equation}
By applying the inequality (\ref{fund5}) for $(\bX,\bY)=(\clHi_{1,k}\bV_k,\Psii_{1,k}(\bV))$ and
$(\bar{X},\bar{Y})=(\Xiota_{1,k},\Yi_{1,k})\triangleq (\clHi_{1,k}\Vi_k,\Psii_{1,k}(\Vi))$,
the following tight concave quadratic minorant of  $\riota_{1,k}(\bV)$ at $\Vi$ is obtained:
\begin{align}
\tri_{1,k}(\bV)\triangleq&\ai_{1,k}+ 2\Re\{\la (\Xiota_{1,k})^H(\Yi_{1,k})^{-1}\clHi_{1,k}\bV_k\ra\}\nonumber\\
&-\la \Upsiloni_{1,k}, \sum_{j\in\clK}[\clHi_{1,k}\bV_j]^2+\sigma I_{N_t}\ra\label{pbo3a}\\
=&\ai_{1,k}+2\Re\{\la \clBi_{1,k}\bV_k\ra \}-\la \clCi_{1,k},\sum_{j\in\clK}[\bV_j]^2\ra,\label{pbo3}
\end{align}
with
\begin{equation}\label{pbo3b}
\begin{array}{c}
0\preceq \Upsiloni_{1,k}\triangleq (\Yi_{1,k})^{-1}-(\Yi_{1,k}+[\Xiota_{1,k}]^2)^{-1},\\
\ai_{1,k}\triangleq \riota_{1,k}(\Vi)-\la [\Xiota_{1,k}]^2(\Yi_{1,k})^{-1}\ra
-\sigma\la \Upsiloni_{1,k}\ra,\\
\clBi_{1,k}\triangleq (\Xiota_{1,k})^H(\Yi_{1,k})^{-1}\clHi_{1,k},\\
\clCi_{1,k}\triangleq (\clHi_{1,k})^H\Upsiloni_{1,k}\clHi_{1,k}.
\end{array}
\end{equation}
We thus solve the following nonsmooth convex problem of the computational complexity order of $\clO(K^3N_c^3N_t^3)$
 to generate $\Vio$:
\begin{equation}\label{bo7}
\max_{\bV} \tfi(\bV)\triangleq \min_{k\in\clK}\tri_{1,k}(\bV)\quad\mbox{s.t.}\quad  (\ref{sa6}).
\end{equation}
Note that $f(\Vi,\zi)=\tfi(\Vi)$ and $\tfi(\Vio)>\tfi(\Vi)$ because $\Vio$ is the optimal solution of
(\ref{bo7}), while $f(\Vio,\zi)\geq \tfi(\Vio)$, so we have (\ref{fial1}) as desired.
\subsection{AP alternating optimization}
{\color{black} In the $\iota$-th AP alternating optimization round, we seek the next feasible point $\zio$, while keeping $(\bV,\btheta)$ fixed at $(\Vio,\thetai)$.} To generate $\zio$ so that
\begin{equation}\label{fial2}
F_{\gamma}(\Vio,\zio,\thetai)> F_{\gamma}(\Vio,\zi,\thetai),
\end{equation}
we consider the following problem of  alternating optimization in $\bz$ for (\ref{sa11}) with $(\bV,\btheta)$ held fixed at $(\Vio,\thetai)$:
\begin{equation}\label{ao}
\max_{\bz} F_{\gamma}(\Vio,\bz,\thetai)=\left[f(\Vio,\bz)-\gamma||\bz -e^{\jmath \thetai}||^2\right],
\end{equation}
where by (\ref{chanmode5}) and (\ref{Psiie}),  $f(\Vio,\bz)=\min_{k\in\clK}  \riota_{2,k}(\bz)$ with
\begin{align}\label{phiko4}
\riota_{2,k}(\bz)&\triangleq r_k(\bz, \Vio)\nonumber\\
&={\color{black}\ln\left|I_{N_t}+[\Omegai_{k,k}(\bz)]^2(\Psii_{2,k}(\bz))^{-1}\right|}, k\in\clK,
\end{align}
with
\begin{equation}\label{phiko5}
\begin{array}{c}
\Psii_{2,k}(\bz)\triangleq \sum_{k'\in\clK\setminus\{k\}}[\Omegai_{k,k'}(\bz)]^2+\sigma I_{N_t},\\
\Omegai_{k,k'}(\bz)\triangleq \begin{bmatrix}\Omegai_{k,k',n_t,n'_t}\bz\end{bmatrix}_{(n_t,n'_t)\in\clN_t\times\clN_t}, (k,k')\in\clK\times\clK,\\
\Omegai_{k,k',\ell,\ell'}\triangleq \tilde{H}_{k,\ell,\ell'}(\Vio_{k'}), (\ell,\ell')\in \clN_t\times\clN_t.
\end{array}
\end{equation}
By applying the inequality (\ref{fund5}) for $(\bX,\bY)=(\Omegai_{k,k}(\bz),\Psii_{2,k}(\bz))$ and
$(\bar{X},\bar{Y})=(\Xiota_{2,k},\Yi_{2,k})\triangleq (\Omegai_{k,k}(\zi),\Psii_{2,k}(\zi))$,
the following tight concave quadratic minorant of  $\riota_{2,k}(\bz)$ at $\zi$ is obtained:
\begin{align}
\tri_{2,k}(\bz)\triangleq&\ai_{2,k}+ 2\Re\{\la (\Xiota_{2,k})^H(\Yi_{2,k})^{-1}\Omegai_{k,k}(\bz)\ra\}\nonumber\\
&-\la \Upsiloni_{2,k}, \sum_{k'\in\clK}[\Omegai_{k,k'}(\bz)]^2\ra\label{phi3a}\\
=&\ai_{2,k}+2\Re\{\bi_{2,k}\bz\}-\la \clCi_{2,k},[\bz]^2\ra,\label{phi3}
\end{align}
with
\begin{equation}\label{phi3b}
\begin{array}{c}
0\preceq \Upsiloni_{2,k}\triangleq (\Yi_{2,k})^{-1}-(\Yi_{2,k}+[\Xiota_{2,k}]^2)^{-1},\\
\clCi_{2,k}\triangleq \sum_{k'\in\clK}\sum_{\ell\in\clN_t}\sum_{\ell'\in\clN_t}(\sqrt{\Upsilon_{2,k}}(\ell,\ell')\Omegai_{k,k',\ell',\ell})^H\\
\times(\sqrt{\Upsilon_{2,k}}(\ell,\ell')\Omegai_{k,k',\ell',\ell}),\\
\ai_{2,k}\triangleq \riota_{2,k}(\zi)-\la [\Xiota_{2,k}]^2(\Yi_{2,k})^{-1}\ra-
\sigma\la \Upsiloni_{2,k}\ra,\\
\bi_{2,k}\triangleq \sum_{\ell\in\clN_t}\sum_{\ell'\in\clN_t}\xi_{k,\ell,\ell'}\Omegai_{k,k,\ell',\ell},
\end{array}
\end{equation}
where $\xi_{k,\ell,\ell'}$ are the entries of the matrix $(\Xiota_{2,k})^H(\Yi_{2,k})^{-1}\in\mathbb{C}^{N_t\times N_t}$.

We thus solve the following nonsmooth convex problem of the computational complexity order of $\clO(L_c^3N_c^3)$
to generate $\zio$:
\begin{equation}\label{ao7}
\max_{\bz} \tF_{\gamma}(\bz)\triangleq\left[ \min_{k\in\clK}\tri_{2,k}(\bz)-\gamma||\bz -e^{\jmath \chii}||^2\right].
\end{equation}
Note that $F_{\gamma}(\Vio,\zi,\thetai)=\tF_{\gamma}(\zi)$ and $\tF_{\gamma}(\zio)>\tF_{\gamma}(\zi)$ because $\zio$ is the optimal solution of
(\ref{ao7}), while $F_{\gamma}(\Vio,\zio,\thetai)\geq \tF_{\gamma}(\zio)$, so we have (\ref{fial2}) as desired.
\subsection{Low-resolution alternating optimization}
To generate $\thetaio$ so that
\begin{align}\label{fial3}
&F_{\gamma}(\Vio,\zio,\thetaio)> F_{\gamma}(\Vio,\zio,\thetai)\nonumber\\
\Leftrightarrow& ||\zio-e^{\jmath\thetaio}||^2<||\zio-e^{\jmath\thetai}||^2
\end{align}
we consider the following problem of  alternating optimization in $\btheta$ for (\ref{sa11}) with $(\bV,\bz)$ held fixed at $(\Vio,\zio)$:
\begin{equation}\label{to}
\min_{\btheta} ||\zio -e^{\jmath \btheta}||^2\quad \mbox{s.t.}\quad (\ref{br1}),
\end{equation}
which admits the closed-form solution
\begin{equation}\label{the3a}
\thetaio_{n_c,\ell_c}=\lfloor\angle\zio_{n_c,\ell}\rceil_{b}, (n_c,\ell_c)\in \clN_c\times\clL_c.
\end{equation}
\subsection{Max-min log-det algorithm and its convergence}
Clearly, (\ref{fial}) follows from (\ref{fial1}), (\ref{fial2}), and (\ref{fial3}), i.e. the alternating optimization procedure generates a sequence $\{ \Vi,\zi,\thetai \}$ of progressively improved
 feasible points for (\ref{sa11}) {\color{black} by solving the convex problems (\ref{bo7}), (\ref{ao7}) and (\ref{to})}, ultimately converging to a feasible point
$(\bar{V},\bar{z},\bar{\theta})$ by Cauchy's theorem. Moreover, with the penalty factor $\gamma$ sufficiently large, $\zi-e^{\jmath\thetai}$ converges to zero, {\color{black} indicating that $(\bar{z},\bar{\theta})$ satisfies the nonlinear constraint (\ref{naosa2e}). Thus, $(\bar{V},\bar{z},\bar{\theta})$ is also a feasible point for (\ref{basic}).} For further insights into the optimality of the penalized optimization approach, the reader is referred to
\cite[Chapter 16]{Betal06} and \cite{PTKD12,CTN14}. The pseudo-code for implementing this alternating
optimization procedure is provided by Algorithm \ref{aalg1}.

\begin{algorithm}[t]
	\caption{Nonsmooth max-min throughput optimization algorithm} \label{aalg1}
	\begin{algorithmic}[1]
	\State \textbf{Initialization:} Initialize $(V^{(0)}, z^{(0)}, \theta^{(0)})$ feasible for (\ref{sa11}). Set $\iota=1$.
		\State \textbf{Repeat until convergence:} Generate
$\Vio$ by solving the convex problem  (\ref{bo7}) of the computational complexity $\clO(K^3N_c^3N_t^3)$. Generate $\zio$ by solving the convex problem  (\ref{ao7}) of the computational complexity
$\clO(L_c^3N_c^3)$. Generate $\thetaio$ by the closed-form (\ref{the3a}).
Reset $\iota\leftarrow \iota+1$.
 \State \textbf{Output} $(\Vi, \zi, \thetai)$ and the resultant $r_k(\zi, \Vi)$.
	\end{algorithmic}
\end{algorithm}
\subsection{Smooth sum-throughput  maximization algorithm}
Instead of the max-min log-det optimization problem (\ref{basic}), we now consider the following problem of sum log-det maximization
\begin{equation}\label{sum}
\max_{\bz,\btheta,\bV} \sum_{k\in\clK}r_k(\bV,\bz) \quad\mbox{s.t.}\quad  (\ref{naosa3}), (\ref{naosa2e}), (\ref{br1}), (\ref{sa6}),
\end{equation}
which  is then reformulated to the following penalized optimization problem
\begin{align}\label{mgm}
\max_{\bV,\bz,\btheta} F_{\gamma,S}(\bV,\bz,\btheta)\triangleq \left[\sum_{k\in\clK}  r_k(\bV,\bz)-\gamma||\bz-e^{\jmath\btheta}||^2\right]
\nonumber\\
\mbox{s.t.}\quad (\ref{br1}), (\ref{sa6}),
\end{align}
where  like (\ref{sa11}), $\gamma>0$ is the penalty factor.
Initialized by $(V^{(0)}, z^{(0)}, \theta^{(0)})$ feasible for
(\ref{mgm}), let $(\Vi,\zi,\thetai)$ be a feasible point for (\ref{mgm}) that is found from the $(\iota-1)$-st iteration. To generate $\Vio$ so that
\begin{equation}\label{sum1}
F_{\gamma,S}(\Vio,\zi,\thetai) > F_{\gamma,S}(\Vi,\zi,\thetai)
\end{equation}
we consider the following counterpart of (\ref{bo}):
\begin{equation}\label{bmgm1}
\max_{\bV} \sum_{k\in\clK}\riota_{1,k}(\bV)
\quad\mbox{s.t.}\quad (\ref{sa6}),
\end{equation}
where $\riota_{1,k}(\bV)$ is defined from (\ref{bmgm2}). Using the tight minorant $\tri_{1,k}(\bV)$
of $\riota_{1,k}(\bV)$ defined
from (\ref{pbo3}), a tight minorant of $\sum_{k\in\clK}\riota_{1,k}(\bV)$ is obtained as
\begin{align}
\gi_1(\bV)\triangleq&\sum_{k\in\clK}\tri_{1,k}(\bV)\nonumber\\
=&\sum_{k\in\clK}\ai_{1,k}+2\sum_{k\in\clK}\Re\{\la \clBi_{1,k}\bV_k\ra\}
-\sum_{k\in\clK}\la\clCi_{1},[\bV_{k}]^2\ra,
\label{gmbo9}
\end{align}
for
\begin{equation}\label{gmbo10}
\clCi_{1}\triangleq \sum_{k\in\clK}\clCi_{1,k}.
\end{equation}
We thus solve the following convex quadratic problem {\color{black}of the computational complexity order of $\clO(KN_cN_t)$} to generate $\Vio$ verifying (\ref{sum1}):
\begin{equation}\label{gmbo11}
\max_{\bV} \left[2\sum_{k\in\clK}\Re\{\la \clBi_{1,k}\bV_k\ra\}-\sum_{k\in\clK}\la\clCi_{1},[\bV_{k}]^2\ra \right]\quad\mbox{s.t.}\quad (\ref{sa6}),
\end{equation}
which admits the closed form solution
\begin{equation}\label{gmbo12}
\Vio_k=\begin{cases}\begin{array}{ll}(\clCi_{1})^{-1}(\clBi_{1,k})^H\\ \mbox{if}\
\sum_{k\in\clK}||(\clCi_{1})^{-1}(\clBi_{1,k})^H||^2\leq P_L,\cr
(\clCi_{1}+\mu I_{N_c})^{-1}(\clBi_{1,k})^H\\ \mbox{otherwise},
\end{array}
\end{cases}
\end{equation}
where $\mu>0$ is found by bisection so that $\sum_{k\in\clK}||(\clCi_{1}+\mu I_{N_c})^{-1}(\clBi_{1,k})^H||^2=P_L$.

To generate $\zio$ for ensuring
\begin{equation}\label{sum2}
F_{\gamma,S}(\Vio,\zio,\thetai) > F_{\gamma,S}(\Vio,\zi,\thetai)
\end{equation}
we consider  the following counterpart of (\ref{ao}):
\begin{equation}\label{amgm1}
\max_{\bz} \sum_{k\in\clK}\riota_{2,k}(\bz)-\gamma||\bz -e^{\jmath \thetai}||^2,
\end{equation}
with $\riota_{2,k}(\bz)$ defined from (\ref{phiko4})-(\ref{phiko5}).
Using the tight minorant $\tri_{2,k}(\bz)$
of $\riota_{2,k}(\bz)$ defined
from (\ref{phi3}), a tight minorant of $\sum_{k\in\clK}\riota_{2,k}(\bz)$ is obtained as
\begin{align}
\gi_2(\bz)&\triangleq\sum_{k\in\clK}\tri_{2,k}(\bz)\nonumber\\
&=\sum_{k\in\clK}\ai_{2,k}+2\Re\{ \bi_{2}\bz \}-\la \clCi_{2},[\bz]^2\ra,
\label{amgm3}
\end{align}
for
\begin{equation}\label{amgm4}
\bi_{2}\triangleq \sum_{k\in\clK}\bi_{2,k}\quad\&\quad
\clCi_{2}\triangleq \sum_{k\in\clK}\clCi_{2,k}.
\end{equation}
We thus solve the following convex quadratic problem {\color{black}of the computational complexity order of $\clO(L_cN_c)$} to generate $\zio$ verifying (\ref{sum2}):
\begin{equation}\label{ampo11}
\max_{\bz}\left[ 2\Re\{\bi_{2}\bz \}-\la \clCi_{2},[\bz]^2\ra-\gamma||\bz -e^{\jmath \thetai}||^2
\right],
\end{equation}
which admits the closed form solution
\begin{equation}\label{ampo12}
\zio=(\clCi_{2}+\gamma I_{\color{black}N_cL_c})^{-1}\left((\bi_2)^H+\gamma e^{\jmath \thetai} \right).
\end{equation}
Lastly, using the closed-form expression of $\thetaio$ in (\ref{the3a}), it becomes clear that
\begin{equation}\label{sum3}
F_{\gamma,S}(\Vio,\zio,\thetaio) > F_{\gamma,S}(\Vio,\zio,\thetai).
\end{equation}
The pseudo-code for implementing the alternating
optimization procedure based on (\ref{gmbo12}), (\ref{ampo12}), and (\ref{the3a})
is provided by Algorithm \ref{aalg2}. It follows from (\ref{sum1}), (\ref{sum2}), and (\ref{sum3}) that
$F_{\gamma,S}(\Vio,\zio,\thetaio) > F_{\gamma,S}(\Vi,\zi,\thetai)$, which
ensures the convergence of Algorithm \ref{aalg2}.

\begin{algorithm}[t]
	\caption{Scalable-complexity sum-throughput maximization algorithm} \label{aalg2}
	\begin{algorithmic}[1]
	\State \textbf{Initialization:} Initialize $( V^{(0)}, z^{(0)}, \theta^{(0)})$ feasible for (\ref{mgm}). Set $\iota=1$.
		\State \textbf{Repeat until  convergence:} Generate
$\Vio$ by the closed-form (\ref{gmbo12}) {\color{black}of the computational complexity $\clO(KN_cN_t)$}.  Generate
$\zio$ by the closed-form (\ref{ampo12}) {\color{black}of the computational complexity $\clO(L_cN_c)$}. Generate $\thetaio$ by the closed-form (\ref{the3a}).
Reset $\iota\leftarrow \iota+1$.
 \State \textbf{Output} $(\Vi, \zi, \thetai)$, the resultant rates $r_k(\Vi,\zi)$ and their sum.
	\end{algorithmic}
\end{algorithm}
\section{Soft max-min optimization for Pareto optimization}
Algorithm \ref{aalg2} conceived  for solving the sum log-det maximization problem (\ref{sum}) is much more computationally efficient and thus practical than Algorithm \ref{aalg1} conceived  for the max-min log-det optimization problem  (\ref{basic}). This is because the former iterates the convex quadratic problems (\ref{gmbo11}) and (\ref{ampo11}), which admit the closed form solutions (\ref{gmbo12}) and
(\ref{ampo12}) with scalable complexity while the latter iterates the nonsmooth  convex problems (\ref{bo7}) and (\ref{ao7}) associated  with cubic complexities. However,
the sum log-det is maximized by differentiating the individual log-det  values  and
thus it is not capable of ensuring the target QoD for all users. In this section, following
\cite{Tuaetal23}, we scale up the log-det function $r_k(\bV,\bz)$ defined from
(\ref{chanmode4})-(\ref{chanmode5}) as
\begin{equation}\label{pa1}
r_{k,\delta}(\bV,\bz)\triangleq {\color{black}\ln\left|I_{N_t}+\frac{1}{\delta}[\clH_k(\bz)\bV_k]^H
\Psi_k^{-1}(\bV,\bz)[\clH_k(\bz)\bV_k]\right|}
\end{equation}
with $0<\delta\leq 1$. For $f_{\delta}(\bV,\bz)\triangleq \min_{k\in\clK}r_{k,\delta}(\bV,\bz)$,
it may be seen that $f_{\delta}(\bV,\bz)> f(\bV,\bz)$ for
$0<\delta<1$, and $\max_{\bV,\bz}f_{\delta}(\bV,\bz)\Leftrightarrow \max_{\bV,\bz}f(\bV,\bz)$ for $N_t=1$. Instead of maximizing $f(\bV,\bz)$ in (\ref{basic}), we opt for maximizing the function
$f_{\delta}(\bV,\bz)$, which is still nonsmooth, but admits the two-sided approximation
$f_{\delta}(\bV,\bz)\geq -f_{SA}(\bV,\bz)\geq f_{\delta}(\bV,\bz)-\ln K$,
with $f_{SA}(\bV,\bz)\triangleq {\color{black}\ln \sum_{k\in\clK}\left|I_{N_t}+\frac{1}{\delta}[\clH_k(\bz)\bV_k]^2
\Psi_k^{-1}(\bV,\bz)\right|^{-1}}$, which is a smooth function. Thus, the smooth function $-f_{SA}(\bV,\bz)$ is regarded as a soft-min function \cite{EMMZ20}.
Instead of the problem (\ref{basic}) of  nonsmooth max-min optimization,
we now consider the following problem of smooth soft max-min optimization:
\begin{equation}\label{basicsoft}
\max_{\bz,\btheta,\bV}\ [-f_{SA}(\bV,\bz)] \quad\mbox{s.t.}\quad  (\ref{naosa3}), (\ref{naosa2e}), (\ref{br1}), (\ref{sa6}).
\end{equation}
More importantly, we will demonstrate through simulations  in the next section that our HP based on
(\ref{basicsoft}) is Pareto-optimal, as it finds the min log-det value as effectively as the  HP based on the max min problem (\ref{basic}), while also attaining the sum log-det value competently as the HP based on the sum log-det problem (\ref{sum}).

Observe that the problem (\ref{basicsoft}) is equivalent to the following problem
\begin{equation}\label{bamsm}
\min_{\bV,\bz} \varphi(\bV,\bz)\triangleq \ln\Xi_{\delta}(\bV,\bz)\quad\mbox{s.t.}\quad  (\ref{naosa3}), (\ref{naosa2e}), (\ref{br1}), (\ref{sa6}),
\end{equation}
where
\begin{align}\label{pic}
\Xi_{\delta}(\bV,\bz)\triangleq &\sum_{k\in\clK}\left(I_{N_t}- [\clH_{k}(\bz)\bV_{k}]^H \right.\nonumber\\
&\left.\times\left([\clH_{k}(\bz)\bV_{k}]^2+\delta\Psi_k(\bz, \bV)\right)^{-1}[\clH_{k}(\bz)\bV_{k}]\right).
\end{align}
The penalized optimization reformulation for (\ref{bamsm}) is
\begin{align}\label{msm}
\min_{\bV,\bz} \Phi_{\gamma}(\bV,\bz,\btheta)\triangleq \left[\ln\Xi_{\delta}(\bV,\bz) +\gamma||\bz-e^{\jmath\btheta}||^2 \right]\nonumber\\
\mbox{s.t.}\quad (\ref{br1c}), (\ref{sa6}),
\end{align}
where $\gamma>0$ is the penalty factor.
Initialized by  $( V^{(0)}, z^{(0)}, \theta^{(0)})$ feasible for
(\ref{msm}), let $(\Vi, \zi, \thetai)$ be a feasible point for (\ref{msm}) that is found from the $(\iota-1)$-st iteration. The alternating optimization procedure at the $\iota$-th-th iteration to generate $(\Vio, \zio, \thetaio)$ feasible for (\ref{msm})
unfolds as follows.
\subsection{DP alternating optimization}
Like (\ref{bo}) and (\ref{bmgm1}), we consider the following problem to generate $\Vio$:
\begin{equation}\label{bsm1}
\min_{\bV} \varphi(\bV,\zi)=\ln \Xiii_{1,\delta}(\bV)
\quad\mbox{s.t.}\quad (\ref{sa6})
\end{equation}
where
\begin{align}\label{bsm1a}
\Xiii_{1,\delta}(\bV)\triangleq& \Xi_{\delta}(\bV,\zi)\nonumber\\
=&\sum_{k\in\clK}\left(I_{N_t}- [\clHi_{1,k}\bV_{k}]^H\right.\nonumber\\
&\left.\times\left([\clHi_{1,k}\bV_{k}]^2
+\delta\Psii_{1,k}(\bV)\right)^{-1}[\clHi_{1,k}\bV_{k}]\right)
\end{align}
with $\Psii_{1,k}(\bV)$ defined from (\ref{bo4}) and $\clHi_{1,k}$ defined from (\ref{bo5}).

Applying the inequality (\ref{ap6}) for $(\bX_k,\bY_k)=(\clHi_{1,k}\bV_k, [\clHi_{1,k}\bV_k]^2
		+\delta\Psii_{1,k}(\bV))$,
$k\in\clK$, and $(\bar{X}_k,\bar{Y}_k)=(\Xiota_{1,k},\Yi_{1,k})\triangleq (\clHi_{1,k}\Vi_k, [\clHi_{1,k}\Vi_k]^2 +\delta\Psii_{1,k}(\Vi))$,
yields the following tight majorant of $\ln \Xiii_{1,\delta}(\bV)$ at $\Vi$:
\begin{align}
\tvarphii_{1}(\bV)\triangleq& \ai_1-2\sum_{k\in\clK}\Re\{\la\clBi_{1,k}\bV_k\ra\}\nonumber\\
&+\sum_{k\in\clK}\la\tclCi_{1,k},[\clHi_{1,k}\bV_k]^2
+\delta\sum_{j\in\clK\setminus\{k\}}[\clHi_{1,k}\bV_j]^2\ra\label{bsm6}\\
=&\ai_1-2\sum_{k\in\clK}\Re\{\la\clBi_{1,k}\bV_k\ra\}+
\sum_{k\in\clK}\la \clCi_{1,k},[\bV_k]^2\ra, \label{bsm8}
\end{align}
where we have (\ref{bsm8}) shown at the top of next page.
\begin{figure*}[t]
\begin{equation}\label{bsm8}
\begin{array}{lll}
	\ai_1&\triangleq& \ln \Xiii_{1,\delta}(\Vi)+ \sum_{k\in\clK}\la [\Xiii_{1,\delta}(\Vi)]^{-1}(\Xiota_{1,k})^H(\Yi_{1,k})^{-1}\Xiota_{1,k}\ra+\delta\sigma\sum_{k\in\clK}\la \tclCi_{1,k}\ra,\\ 	
	\clBi_{1,k}&\triangleq& [\Xiii_{1,\delta}(\Vi)]^{-1}(\Xiota_{1,k})^H(\Yi_{1,k})^{-1}\clHi_{1,k},\\
	\tclCi_{1,k}&\triangleq& (\Yi_{1,k})^{-1}\Xiota_{1,k}[\Xiii_{1,\delta}(\Vi)]^{-1}(\Xiota_{1,k})^H(\Yi_{1,k})^{-1}, k\in\clK,\\
	\clCi_{1,k}&\triangleq& (\clHi_{1,k})^H \tclCi_{1,k}\clHi_{1,k}+
	\delta\sum_{j\in\clK\setminus\{k\}}(\clHi_{1,j})^H \tclCi_{1,j}\clHi_{1,j}, k\in\clK.
\end{array}
\end{equation}
\hrulefill
\vspace{-0.2cm}
\end{figure*}
We thus solve the following problem of tight majorant minimization {\color{black}of the computational complexity order of $\clO(KN_cN_t)$} to generate $\Vio$:
\begin{equation}\label{smbo11}
\min_{\bV} \tvarphii_{1}(\bV)\quad\mbox{s.t.}\quad (\ref{sa6}),
\end{equation}
which admits the closed-form solution
\begin{equation}\label{smbo12}
\Vio_k=\begin{cases}\begin{array}{ll}
\!(\clCi_{1,k})^{-1}(\clBi_{1,k})^H\\ \!\mbox{if}\ \sum_{k\in\clK}||(\clCi_{1,k})^{-1}(\clBi_{1,k})^H||^2\leq P_L,\cr
\!(\clCi_{1,k}+\mu I_{N_c})^{-1}(\clBi_{1,k})^H\\ \!\mbox{otherwise},
\end{array}
\end{cases}
\end{equation}
where $\mu>0$ is found by bisection so that $\sum_{k\in\clK}||(\clCi_{1,k}+\mu I_{N_c})^{-1}(\clBi_{1,k})^H||^2=P_L$. Like in (\ref{fial1}), we have
\begin{equation}\label{sfial1}
\Phi_{\gamma}(\Vio,\zi,\thetai)< \Phi_{\gamma}(\Vi,\zi,\thetai).
\end{equation}
\subsection{AP  alternating optimization}
Similarly to  (\ref{ao}) and (\ref{amgm1}),
we consider the following problem to generate $\zio${
\begin{equation}\label{asm1}
\min_{\bz} \Phi_{\gamma}(\Vio,\bz,\thetai)=\left[\ln \Xiii_{2,\delta}(\bz)+\gamma||\bz -e^{\jmath \thetai}||^2\right],
\end{equation}
where we have
\begin{align}\label{asm1a}
\Xiii_{2,\delta}(\bz)\triangleq&\Xi_{\delta}(\Vio,\bz)\nonumber\\
=&\sum_{k\in\clK}\left(I_{N_t}- [\Omegai_{k,k}(\bz)]^H\right.\nonumber\\
&\left.\times\left([\Omegai_{k,k}(\bz)]^2
+\delta\Psii_{2,k}(\bz)\right)^{-1}[\Omegai_{k,k}(\bz)]\right)
\end{align}
with $\Psii_{2,k}(\bz)$ and $\Omegai_{k,k}(\bz)$ defined from (\ref{phiko5}).

Using the inequality (\ref{ap6}) for $(\bX_k,\bY_k)=(\Omegai_{k,k}(\bz),
[\Omegai_{k,k}(\bz)]^2+\delta\Psii_{2,k}(\bz))$,
 and $(\bar{X}_k,\bar{Y}_k)=(\Xiota_{2,k},\Yi_{2,k})\triangleq (\Omegai_{k,k}(\zi), [\Omegai_{k,k}(\zi)]^2 +\delta\Psii_{2,k}(\zi))$, $k\in\clK$,
yields the following tight majorant of $\ln \Xiii_{2,\delta}(\bz)$ at $\zi$:
\begin{align}
\tvarphii_{2}(\bz)\triangleq& \ai_2-2\sum_{k\in\clK}\Re\{\la\clBi_{2,k}\Omegai_{k,k}(\bz)\ra\}\nonumber\\
&+\sum_{k\in\clK}\la\Upsiloni_{2,k},[\Omegai_{k,k}(\bz)]^2
+\delta\sum_{k'\in\clK\setminus\{k\}}[\Omegai_{k,k'}(\bz)]^2\ra\label{asm6}\\
=&\ai_2-2\sum_{k\in\clK}\Re\{\bi_{2,k}\bz\}+
\sum_{k\in\clK}\la \clCi_{2,k},[\bz]^2\ra \label{asm8}\\
=&\ai_2-2\Re\{\bi_{2}\bz\}+
\la \clCi_{2},[\bz]^2\ra, \label{asm8}
\end{align}
where we have (\ref{asm8b}) shown at the top of next page.
\begin{figure*}[t]
\begin{equation}\label{asm8b}
\begin{array}{lll}
	\ai_2&\triangleq &\ln \Xiii_{2,\delta}(\zi)+ \sum_{k\in\clK}\la [\Xiii_{2,\delta}(\zi)]^{-1}(\Xiota_{2,k})^H(\Yi_{2,k})^{-1}\Xiota_{2,k}\ra\\	
	&&+\delta\sum_{k\in\clK}\sigmai_{2,k}\la \Upsiloni_{2,k}\ra,\\
	\clBi_{2,k}&\triangleq& [\Xiii_{2,\delta}(\zi)]^{-1}(\Xiota_{2,k})^H(\Yi_{2,k})^{-1},\\
	\Upsiloni_{2,k}&\triangleq & (\Yi_{2,k})^{-1}\Xiota_{2,k}[\Xiii_{2,\delta}(\zi)]^{-1}(\Xiota_{2,k})^H(\Yi_{2,k})^{-1}, k\in\clK,\\
	\bi_{2,k}&\triangleq & \sum_{\ell\in\clN_t}\sum_{\ell'\in\clN_t}\xi_{k,\ell,\ell'}\Omegai_{k,k,\ell',\ell},\\
	\clCi_{2,k}&\triangleq & \sum_{\ell\in\clN_t}\sum_{\ell'\in\clN_t}(\sqrt{\Upsilon_{2,k}}(\ell,\ell')\Omegai_{k,k,\ell',\ell})^H
	(\sqrt{\Upsilon_{2,k}}(\ell,\ell')\Omegai_{k,k,\ell',\ell})\\
	&&+\delta\sum_{k'\in\clK\setminus\{k\}}\sum_{\ell\in\clN_t}\sum_{\ell'\in\clN_t}(\sqrt{\Upsilon_{2,k}}(\ell,\ell')\Omegai_{k,k',\ell',\ell})^H
	(\sqrt{\Upsilon_{2,k}}(\ell,\ell')\Omegai_{k,k',\ell',\ell}),\\
	\bi_{2}&\triangleq& \sum_{k\in\clK}\bi_{2,k},\\
	\clCi_{2}&\triangleq &\sum_{k\in\clK}\clCi_{2,k},
\end{array}
\end{equation}
where $\xi_{k,\ell,\ell'}$ are the entries of $\clBi_{2,k}\in\mathbb{C}^{N_t\times N_t}$.

\hrulefill
\vspace{-0.4cm}
\end{figure*}

We thus solve the following problem of tight majorant minization {\color{black}of the computational complexity order of $\clO(L_cN_c)$} to generate $\zio$:
\begin{equation}\label{osm11}
\min_{\bz} \left[\tvarphii_{2}(\bz) +\gamma||\bz -e^{\jmath \thetai}||^2\right],
\end{equation}
which admits the closed-form solution
\begin{equation}\label{osm12}
\zio=(\clCi_{2,k}+\gamma I_{\color{black}N_cL_c})^{-1}\left((\bi_2)^H+\gamma e^{\jmath \thetai} \right).
\end{equation}
Similarly to (\ref{fial2}), we have
\begin{equation}\label{sfial2}
\Phi_{\gamma}(\Vio,\zio,\thetai)< \Phi_{\gamma}(\Vio,\zi,\thetai).
\end{equation}
\begin{algorithm}[t]
	\caption{Scalable-complexity soft max-min throughput optimization algorithm} \label{aalg3}
	\begin{algorithmic}[1]
		\State \textbf{Initialization:} Initialize $(z^{(0)}, V^{(0)}, \theta^{(0)})$ feasible for (\ref{msm}). Set $\iota=1$.
		\State \textbf{Repeat until convergence:} Generate
		$\Vio$ by  (\ref{smbo12}) {\color{black}of the computational complexity $\clO(KN_cN_t)$}.  Generate
		$\zio$ by (\ref{osm12}) {\color{black}of the computational complexity $\clO(L_cN_c)$}. Generate $\thetaio$ by (\ref{the3a}).
		Reset $\iota\leftarrow \iota+1$.
		\State \textbf{Output} $(\Vi, \zi, \thetai)$ and the resultant rates.
	\end{algorithmic}
\end{algorithm}
\subsection{Soft max-min algorithm and convergence}
Lastly, upon using $\thetaio$ generated  by (\ref{the3a}), it is seen that
\begin{equation}\label{sfial3}
\Phi_{\gamma,S}(\Vio,\zio,\thetaio) < \Phi_{\gamma,S}(\Vio,\zio,\thetai).
\end{equation}
The pseudo-code of implementing the alternating
optimization procedure {\color{black} for solving problem (\ref{msm})} based on the closed forms (\ref{smbo12}), (\ref{osm12}), and (\ref{the3a})
is provided by Algorithm \ref{aalg3}. It follows from (\ref{sfial1}), (\ref{sfial2}), and (\ref{sfial3}) that
$\Phi_{\gamma}(\Vio,\zio,\thetaio) < \Phi_{\gamma}(\Vi,\zi,\thetai)$, which
ensures the convergence of Algorithm \ref{aalg3}. {\color{black} The sequence $\{ \Vi,\zi,\thetai \}$ of improved feasible points for (\ref{msm}) converges to $(\bar{V},\bar{z},\bar{\theta})$, which is a feasible point for (\ref{bamsm}).}

\section{Numerical Results}\label{sec:simulation}
This section evaluates the performance of our proposed algorithms in the scenario of a $12\times 12$-element uniform circular cylindrical array (UCyA) at the BS and $8$ UEs randomly distributed within a cell having $200$-meter radius. The mmWave channel $H_k\in \mathbb{C}^{N_t \times N}$ connecting  the BS to UE $k$ is modelled as $H_k=\sqrt{\frac{NN_t}{N_{cl} N_{sc}}} \sqrt{10^{-\rho_k/10}} \sum_{c=1}^{N_{cl}} \sum_{\ell = 1}^{N_{sc}} \alpha_{k,c,\ell}\\ a_r\left(\phi_{k,c,\ell}^r\right) a_t^H \left(\phi_{k,c,\ell}^t,\theta_{k,c,\ell}^t \right)$. The path-loss $\rho_k$ for the BS-to-UE $k$ link at distance $d_k$ is expressed as $36.72+35.3\log 10(d_k)$ (in dB). The complex gain $\alpha_{k,c,\ell}$ follows Rayleigh fading. The azimuth angle of departure (arrival, resp.) $\phi_{k,c,\ell}^t$ ($\phi_{k,c,\ell}^r$, resp.), as well as the elevation angle of departure $\theta_{k,c,\ell}^t$ are generated using the Laplacian distribution with random mean cluster angles in the interval $[0, 2\pi)$ and a spread of $10$ degrees for each cluster, with the number of clusters $N_{cl}$ set to $5$ and the number $N_{sc}$ of scatters within each cluster  set to $10$ according to \cite{Akd14}.
Furthermore, the transmit antenna array response vector is expressed as $a_t \left(\phi_{k,c,\ell}^t, \theta_{k,c,\ell}^t \right) = a_t^a \left(\phi_{k,c,\ell}^t, \theta_{k,c,\ell}^t \right) \otimes a_t^e \left(\theta_{k,c,\ell}^t \right)$, where the azimuth and elevation components of the antenna array response,  denoted as $a_t^a \left(\phi_{k,c,\ell}^t, \theta_{k,c,\ell}^t \right)$ and $a_t^e \left(\theta_{k,c,\ell}^t \right)$, respectively, are defined as follows:
\begin{equation}
\begin{array}{c}
	\begin{aligned}
	a_t^a \left(\phi_{k,c,\ell}^t, \theta_{k,c,\ell}^t \right) = &\frac{1}{\sqrt{N_a}} \left[ e^{j\frac{2\pi}{\lambda}r\sin(\theta_{k,c,\ell}^t)\cos(\phi_{k,c,\ell}^t-\varphi_1)},\right.\\
	&\left.\cdots, e^{j\frac{2\pi}{\lambda}r\sin(\theta_{k,c,\ell}^t)\cos(\phi_{k,c,\ell}^t-\varphi_{N_a})}\right]^T,\\
	a_t^e \left(\theta_{k,c,\ell}^t \right) = &\frac{1}{\sqrt{N_e}} \left[ 1,e^{-j\frac{2\pi}{\lambda}h\cos(\theta_{k,c,\ell}^t)} ,\right.\\
	&\left.\cdots, e^{-j\frac{2\pi}{\lambda}h(N_e-1)\cos(\theta_{k,c,\ell}^t)}\right]^T.
	\end{aligned}
\end{array}
\end{equation}
We set the radius of UCyA to $r = 2\lambda$, and the vertical spacing between the adjacent uniform circular arrays (UCAs) to $h = 0.5\lambda$ according to \cite{Linetal21}, with $\lambda$ representing the wavelength. Furthermore, $\varphi_{n_a} = 2\pi(n_a-1)/N_a$ represents the angular difference between the central angle of the $n_a$-th antenna and the first antenna within each UCA. Additionally, the receive antenna array response vector is given by
\begin{equation}
	a_r\! \left(\phi_{k,c,\ell}^r \right)\! = \!\frac{1}{\sqrt{N_t}}\! \left[ 1,e^{j\pi\sin(\phi_{k,c,\ell}^r)} ,
	\cdots, e^{j\pi (N_t-1)\sin(\phi_{k,c,\ell}^r)}\right]^T\!\!,
\end{equation}
under the assumption that each UE uses a uniform linear array (ULA) with antennas spaced at half-wavelength intervals.

The number of RF chains is set to $8$ {\color{black}unless otherwise specified}. The noise power density is set to $-174$ dBm/Hz. For practical implementation, we employ a $b=3$-bit resolution phase shifters. The algorithms terminate when the penalty term value drops below $10^{-1}$.

We use the following legends to specify the proposed implementations:
\begin{itemize}
	\item For the conventional AoSA AP associated with $A=I_L$ in (\ref{naosa1}),
  ``{\sf AoSA-MM}'' refers to the
nonsmooth max-min throughput optimization Algorithm \ref{aalg1};``{\sf AoSA-SMM}'' refers to the scalable-complexity soft max-min throughput optimization Algorithm \ref{aalg3}; ``{\sf AoSA-ST}'' refers to the scalable-complexity sum-throughput maximization Algorithm \ref{aalg2};
	\item For the nAoSA AP with $A$ in (\ref{naosa1}) defined by (\ref{cla1}),
 ``{\sf nAoSA-MM}'' refers to the nonsmooth max-min throughput optimization Algorithm \ref{aalg1};``{\sf nAoSA-SMM}'' refers to the scalable-complexity soft max-min throughput optimization
 Algorithm \ref{aalg3}; ``{\sf nAoSA-ST}'' refers to the scalable-complexity sum-throughput maximization Algorithm \ref{aalg2}.
\end{itemize}

\subsection{Algorithmic convergence}
{\color{black} Selecting an appropriate penalty factor $\gamma$ is crucial for ensuring the convergence of our proposed algorithms. A high value of $\gamma$ may lead to premature termination of the iterations, while a low value of $\gamma$ could slow down convergence.} To achieve a satisfactory convergence speed along with the penalty factor $\gamma$, we commence with a $\gamma$ so that the magnitude of the penalty term aligns with that of the objective. Then
in the subsequent iterations, we gradually increase the value of $\gamma$.
For illustration, when considering the penalty parameter $\gamma$ for implementing
Algorithm \ref{aalg1}, we generate $z^{(0)}$ with the modulus of its entries lower than $1$, and $V^{(0)}$ satisfying the power constraint  (\ref{sa6}). Then
the triplet $(V^{(0)}, z^{(0)}, \theta^{(0)})$ associated with $\theta^{(0)}_{n_c,\ell_c}=\lfloor\angle z^{(0)}_{n_c,\ell}\rceil_{b}, (n_c,\ell_c)\in \clN_c\times\clL_c$ (refer to (\ref{the3a})) presents a feasible point for the problem (\ref{sa11}). For implementing the first iteration,
we set $\gamma=\min_{k\in\clK}r_k(V^{(0)},z^{(0)})/||z^{(0)}-e^{\jmath \theta^{(0)}}||^2$, ensuring that
the objective function $ \min_{k\in\clK}r_k(V^{(0)},z^{(0)})$ corresponds in magnitude to the penalty term $\gamma ||z^{(0)}-e^{\jmath \theta^{(0)}}||^2$. As the iterative process continues,  we use the update $\gamma\rightarrow 1.2\gamma$, whenever $||\zio-e^{\jmath \thetaio}||^2>0.9||\zi-e^{\jmath \thetai}||^2$. {\color{black} This procedure can lead to a gradual reduction of the penalty term to zero and facilitate the convergence of the objective function.}

Fig. \ref{fig:nAoSA_conv_Nt2} characterizes the convergence performance of the proposed algorithms when using $80$ phase shifters in nAoSA for $N_t = 2$ at $P = 100$ mW. To provide a more concise depiction of the convergence behaviors of all the proposed algorithms, we use the mean throughput value within the objective function for sum-throughput optimization, as shown in Fig. \ref{fig:nAoSA_conv_obj_fun_Nt2}. The convergence  of the proposed algorithms recorded for $N_t = 1$ exhibits a similar pattern but a faster convergence due to the reduced number of decision variables.

Table \ref{table:nAoSA_AoSA_min_rate_vs_c} demonstrates the impact of adjusting the coefficient $\delta$ in the scalable-complexity soft max-min optimization Algorithm \ref{aalg3} on the  users' minimum throughput achieved with the aid of $80$ phase shifters in nAoSA and $P = 100$ mW, where $\delta = 0.5$ yields the highest users' minimum throughput, which is then used for the next simulations.

\begin{table*}[!t]
	\centering
	\caption{The users' minimum throughput (bps/Hz) vs. $\delta$ for the soft max-min optimization algorithms achieved by nAoSA and AoSA at $P = 100$ mW.}
	\begin{tabular}{|l|c|c|c|c|c|}
		\hline
		& $\delta = 1$ & $\delta = 0.5$ & $\delta = 0.1$ & $\delta = 0.05$ & $\delta = 0.01$ \\ \hline
		{\sf nAoSA-SMM} $(N_t = 1)$ & 2.50    & 2.57      & 2.44      & 2.19       & 1.31      \\ \hline
		{\sf nAoSA-SMM} $(N_t = 2)$ & 2.80    & 2.92      & 2.16      & 1.39      & 1.18       \\ \hline
		{\sf AoSA-SMM} $(N_t = 1)$ & 2.62    & 2.70      & 2.57      & 2.26       & 1.24      \\ \hline
		{\sf AoSA-SMM} $(N_t = 2)$ & 2.77    & 3.02      & 2.16      & 1.41      & 1.19       \\ \hline
	\end{tabular}
	\label{table:nAoSA_AoSA_min_rate_vs_c}
\end{table*}

\begin{figure}[!t]
	\centering
	\begin{subfigure}[c]{0.4\textwidth}
		\centering
		\includegraphics[width=0.92\textwidth]{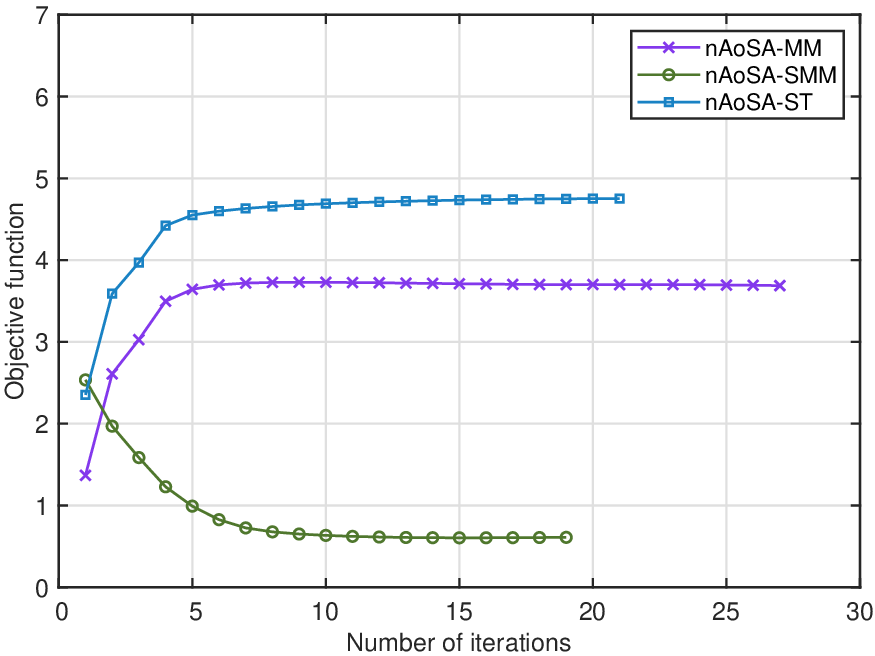}
		\caption{Objective function vs. number of iterations}
		\label{fig:nAoSA_conv_obj_fun_Nt2}
	\end{subfigure}
	\hfill
	\begin{subfigure}[c]{0.4\textwidth}
		\centering
		\includegraphics[width=0.92\textwidth]{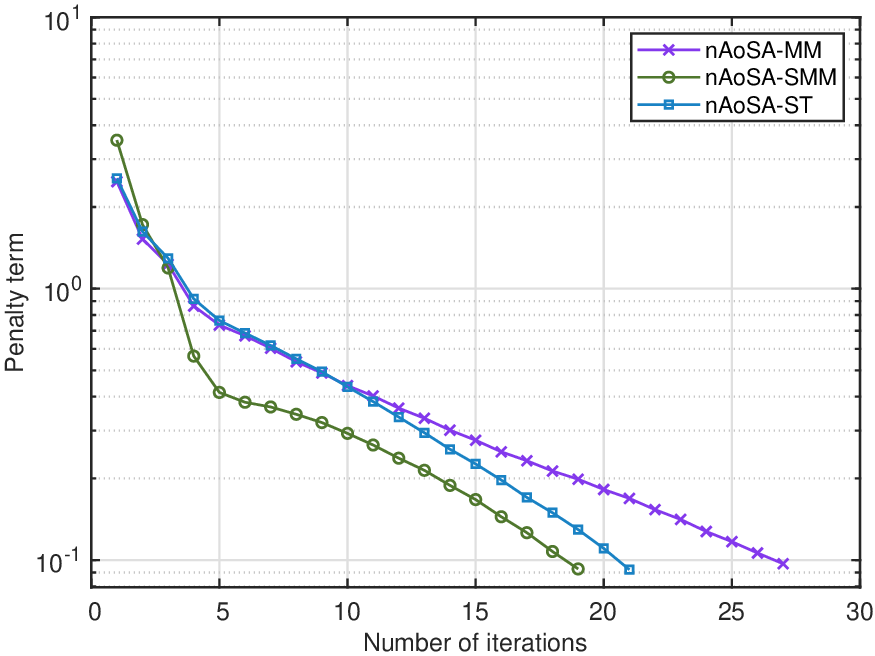}
		\caption{Penalty terms vs. number of iterations}
		\label{fig:nAoSA_conv_penalty_Nt2}
	\end{subfigure}
	\caption{Convergence performance of the proposed algorithms by using nAoSA at $N_t = 2$.}
	\label{fig:nAoSA_conv_Nt2}
\end{figure}

\begin{figure}[!t]
	\centering
	\begin{subfigure}[c]{0.4\textwidth}
		\centering
		\includegraphics[width=0.92\textwidth]{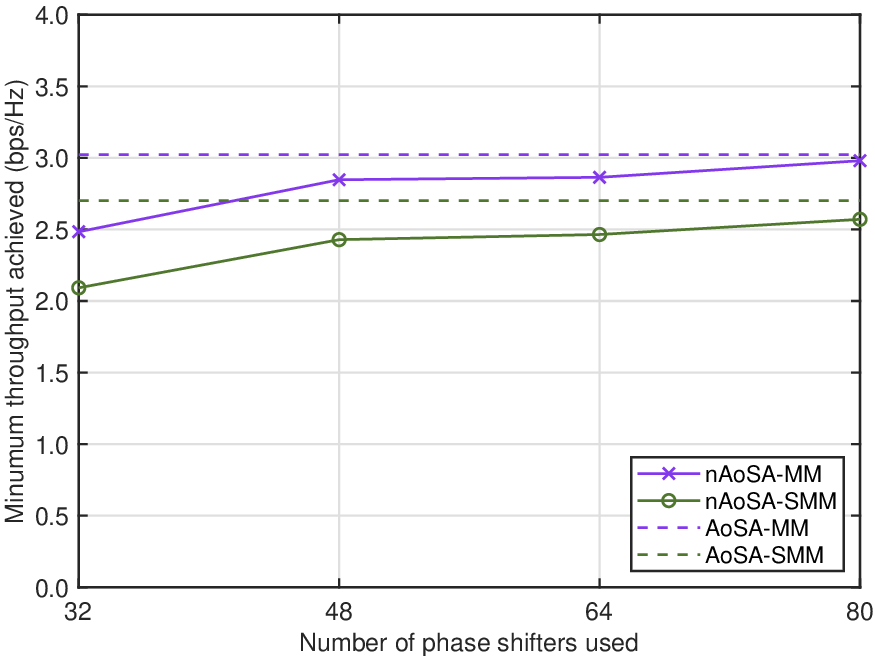}
		\caption{Minimum throughput achieved by the nAoSA under different numbers of phase shifters used vs. the AoSA with $144$ phase shifters {\color{black}and $8$ RF chains}, $N_t = 1$}
		\label{fig:AoSA_nAoSA_P_min_rate_vs_Lc_Nt1}
	\end{subfigure}
	\hfill
	\begin{subfigure}[c]{0.4\textwidth}
		\centering
		\includegraphics[width=0.92\textwidth]{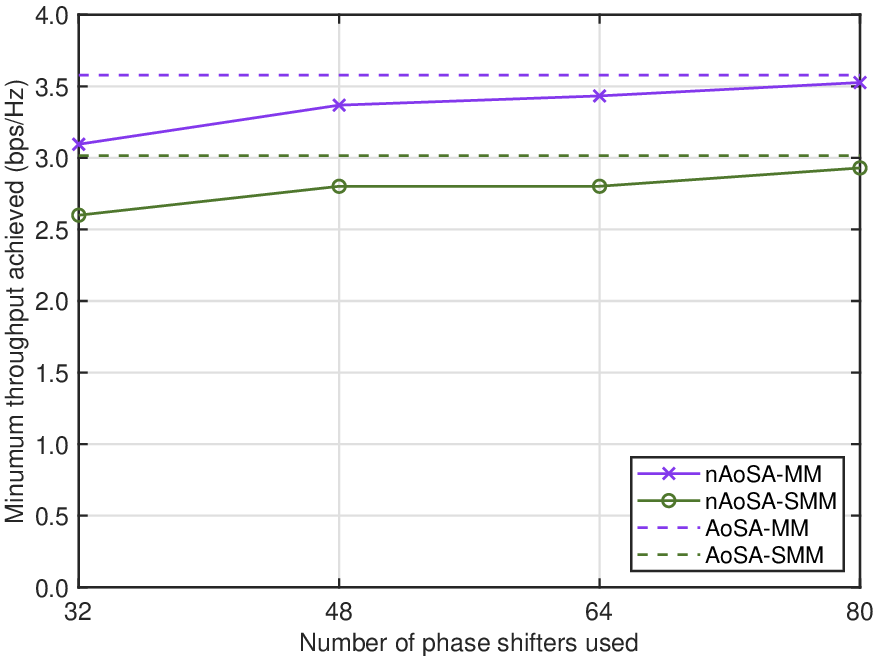}
		\caption{Minimum throughput achieved by the nAoSA under different numbers of phase shifters used vs. the AoSA with $144$ phase shifters {\color{black}and $8$ RF chains}, $N_t = 2$}
		\label{fig:AoSA_nAoSA_P_min_rate_vs_Lc_Nt2}
	\end{subfigure}
	\caption{The minimum throughput achieved by nAoSA vs. the number of phase shifters used at equal transmit power $P$ (number of phase shifters {\color{black}and RF chains} used by AoSA is fixed at $144$ {\color{black}and $8$, respectively}).}
	\label{fig:AoSA_nAoSA_P_min_rate_vs_Lc}
\end{figure}

\begin{figure}[!t]
	\centering
	\begin{subfigure}[c]{0.4\textwidth}
		\centering
		\includegraphics[width=0.92\textwidth]{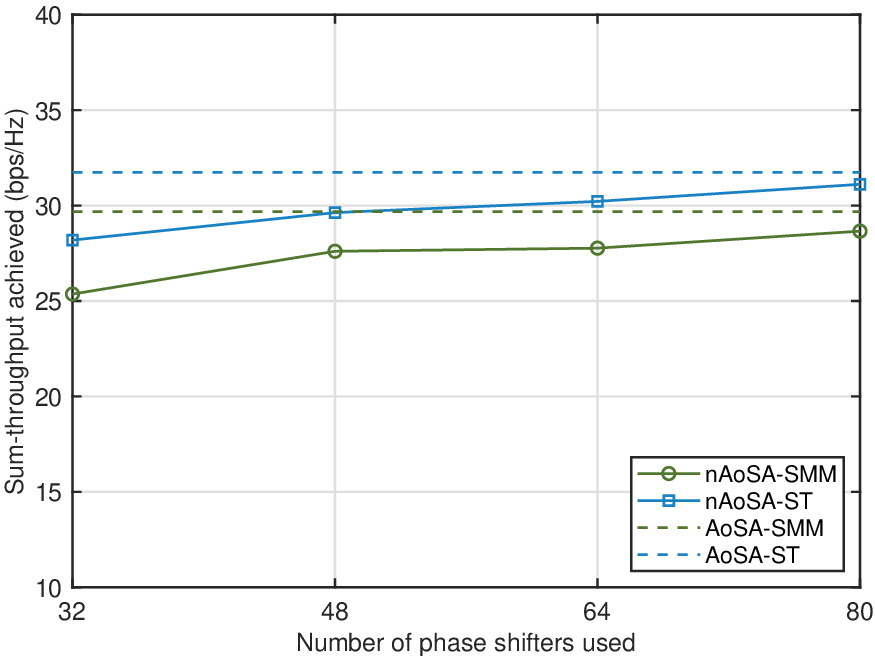}
		\caption{Sum-throughput achieved by the nAoSA under different numbers of  phase shifters used  vs. the AoSA with $144$ phase shifters  {\color{black}and $8$ RF chains}, $N_t = 1$}
		\label{fig:AoSA_nAoSA_P_sum_rate_vs_Lc_Nt1}
	\end{subfigure}
	\hfill
	\begin{subfigure}[c]{0.4\textwidth}
		\centering
		\includegraphics[width=0.92\textwidth]{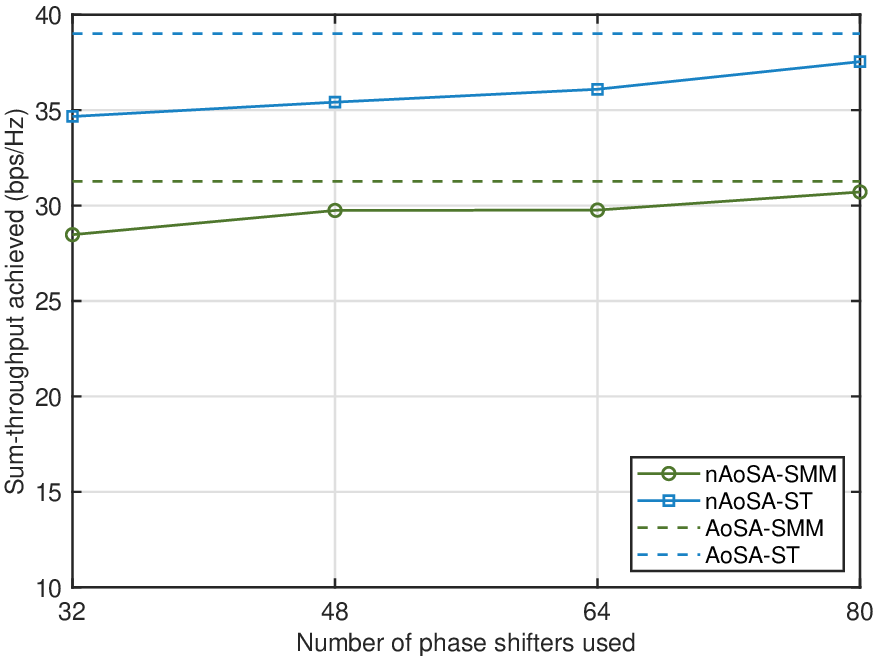}
		\caption{Sum-throughput achieved by the nAoSA under different numbers of phase shifters used vs. the AoSA with $144$ phase shifters {\color{black}and $8$ RF chains}, $N_t = 2$}
		\label{fig:AoSA_nAoSA_P_sum_rate_vs_Lc_Nt2}
	\end{subfigure}
	\caption{The sum-throughput achieved by nAoSA vs. the number of phase shifters used at equal transmit power $P$ (number of phase shifters {\color{black}and RF chains} used by AoSA is fixed at $144$ {\color{black}and $8$, respectively}).}
	\label{fig:AoSA_nAoSA_P_sum_rate_vs_Lc}
\end{figure}

\begin{figure}[!t]
	\centering
	\begin{subfigure}[c]{0.4\textwidth}
		\centering
		\includegraphics[width=0.92\textwidth]{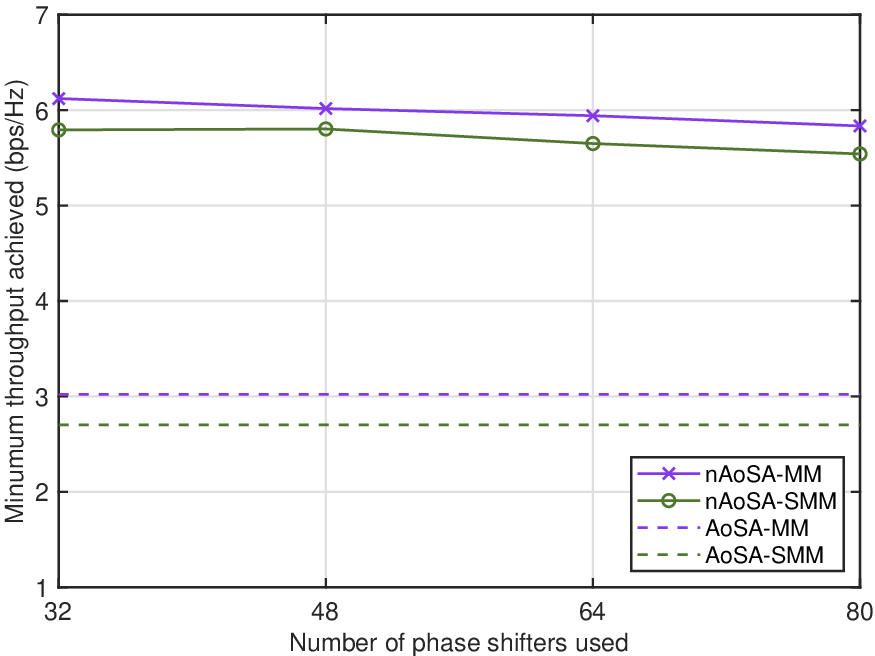}
		\caption{Minimum throughput achieved by the nAoSA under different numbers of phase shifters used  vs. the AoSA with $144$ phase shifters {\color{black}and $8$ RF chains}, $N_t = 1$}
		\label{fig:AoSA_nAoSA_Ptotal_min_rate_vs_Lc_Nt1}
	\end{subfigure}
	\hfill
	\begin{subfigure}[c]{0.4\textwidth}
		\centering
		\includegraphics[width=0.92\textwidth]{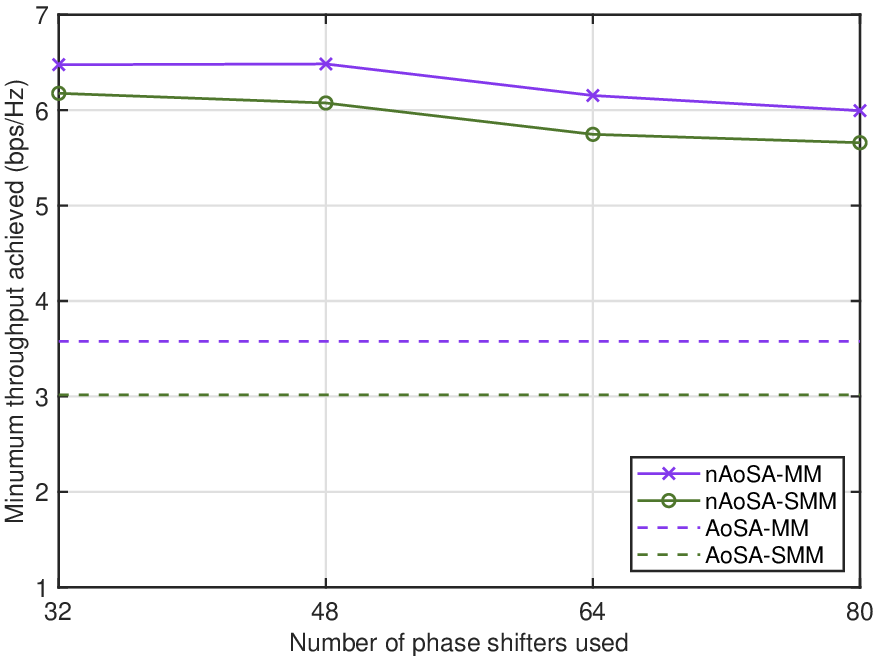}
		\caption{Minimum throughput achieved by the nAoSA under different numbers of phase shifters used  vs. the AoSA with $144$ phase shifters {\color{black}and $8$ RF chains}, $N_t = 2$}
		\label{fig:AoSA_nAoSA_Ptotal_min_rate_vs_Lc_Nt2}
	\end{subfigure}
	\caption{The minimum throughput achieved by nAoSA vs. the number of  phase shifters used
		at equal total power $P_{total}$ (number of phase shifters {\color{black}and RF chains} used by AoSA is fixed at $144$ {\color{black}and $8$, respectively}).}
	\label{fig:AoSA_nAoSA_Ptotal_min_rate_vs_Lc}
\end{figure}

\begin{figure}[!t]
	\centering
	\begin{subfigure}[c]{0.4\textwidth}
		\centering
		\includegraphics[width=0.92\textwidth]{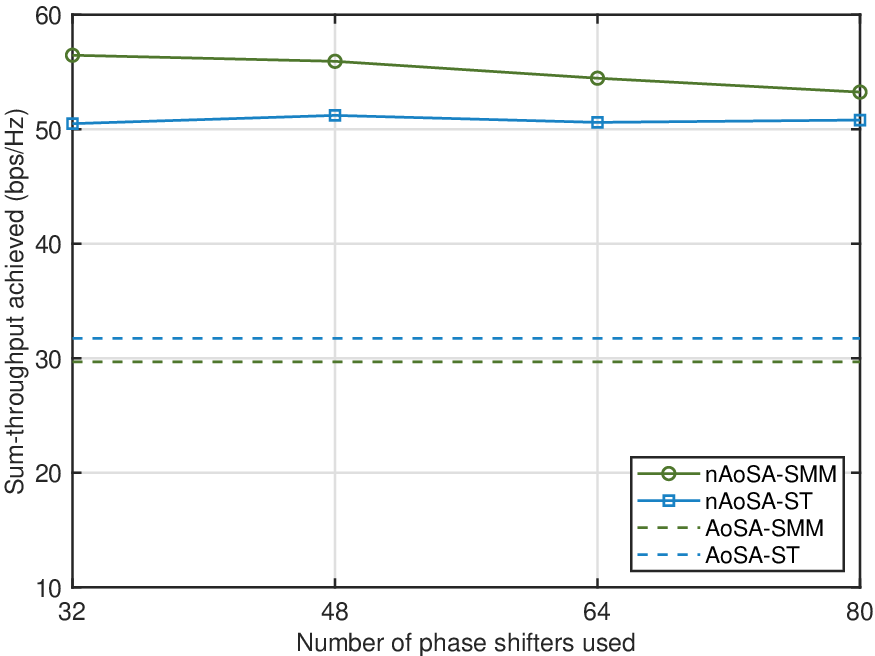}
		\caption{Sum-throughput achieved by the nAoSA under different numbers of phase shifters used  vs. the AoSA with $144$ phase shifters {\color{black}and $8$ RF chains}, $N_t = 1$}
		\label{fig:AoSA_nAoSA_Ptotal_sum_rate_vs_Lc_Nt1}
	\end{subfigure}
	\hfill
	\begin{subfigure}[c]{0.4\textwidth}
		\centering
		\includegraphics[width=0.92\textwidth]{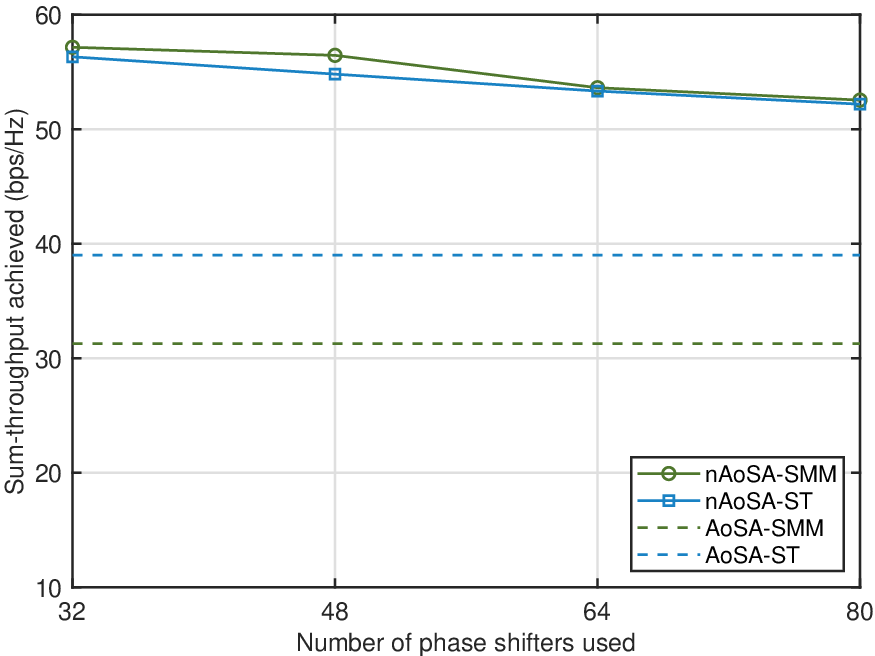}
		\caption{Sum-throughput achieved by the nAoSA under different numbers of phase shifters used  vs. the AoSA with $144$ phase shifters {\color{black}and $8$ RF chains}, $N_t = 2$}
		\label{fig:AoSA_nAoSA_Ptotal_sum_rate_vs_Lc_Nt2}
	\end{subfigure}
	\caption{The sum-throughput achieved by nAoSA vs. the number of  phase shifters used
		at equal total power $P_{total}$ (number of phase shifters {\color{black}and RF chains} used by AoSA is fixed at $144$ {\color{black}and $8$, respectively}).}
	\label{fig:AoSA_nAoSA_nFC_Ptotal_sum_rate_vs_Lc}
\end{figure}

\begin{figure}[!t]
	\centering
	\begin{subfigure}[c]{0.4\textwidth}
		\centering
		\includegraphics[width=0.95\textwidth]{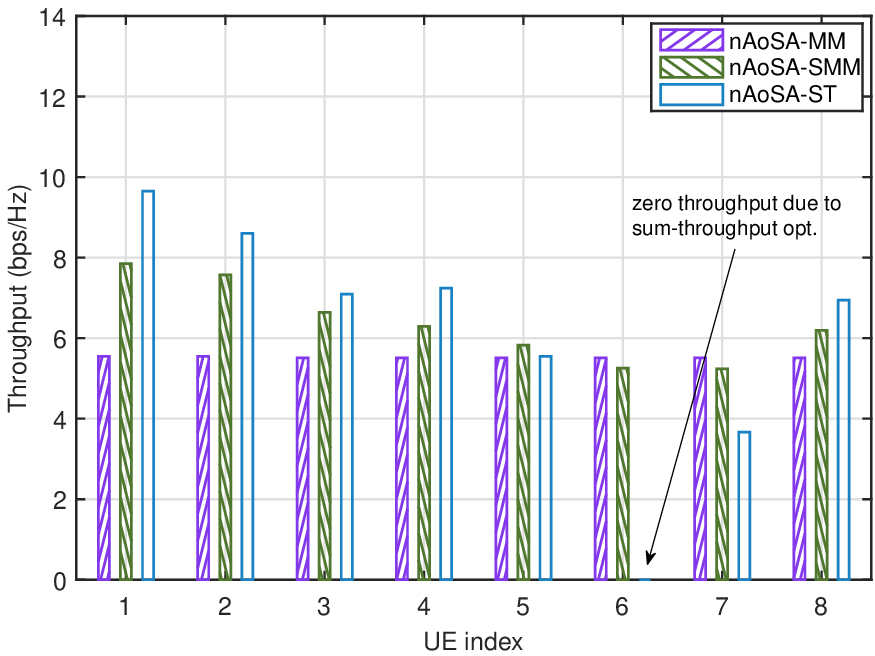}
		\caption{Individual throughputs of all UEs by nAoSA, $N_t = 1$}
		\label{fig:nAoSA_rate_distrib_Lc10_Ptotal_Nt1}
	\end{subfigure}
	\hfill
	\begin{subfigure}[c]{0.4\textwidth}
		\centering
		\includegraphics[width=0.95\textwidth]{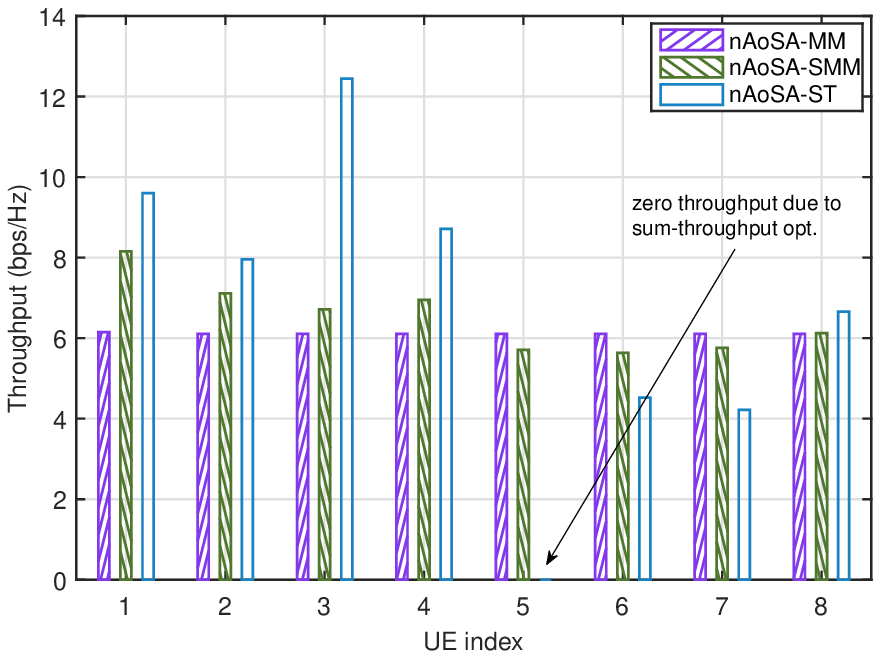}
		\caption{Individual throughputs of all UEs by nAoSA, $N_t = 2$}
		\label{fig:nAoSA_rate_distrib_Lc10_Ptotal_Nt2}
	\end{subfigure}
	\caption{Throughput distributions by nAOSA.}
	\label{fig:nAoSA_rate_distrib_Lc10_Ptotal}
\end{figure}

\subsection{Approaching the AoSA performance by nAoSA at the same transmit power budget}
To start with, we present a comparative analysis of the nAoSA  and the conventional AoSA to demonstrate the following outcomes: $(i)$ Within the same transmit power budget $P$,  nAoSA achieves performances comparable to  AoSA, despite using fewer phase shifters; $(ii)$ The scalable-complexity soft max-min throughput optimization Algorithm \ref{aalg3} attains both excellent users' minimum  throughput and sum-throughput, {\color{black} as it optimizes an approximation of the minimum throughput objective, resulting in a near-optimal solution. Moreover, it achieves significant enhancements in both the minimum throughput and sum-throughput simultaneously, without extensively sacrificing one objective to improve the other.}

Specifically, a comparison between the nAoSA  and  AoSA  in terms of the users' minimum  throughput {\color{black} achieved by the max-min throughput optimization Algorithm \ref{aalg1} and the soft max-min throughput optimization Algorithm \ref{aalg3}} at a transmit power of $100$ mW is presented in Fig. \ref{fig:AoSA_nAoSA_P_min_rate_vs_Lc}. The dashed line represents the minimum user throughput achieved by the AoSA structure {\color{black}employing Algorithm \ref{aalg1}} equipped with {\color{black} an excessive number of} $144$ phase shifters, serving as the reference. {\color{black} When using the max-min throughput optimization Algorithm 1, employing $32$ phase shifters in nAoSA yields approximately $80\%$ of the minimum throughput achieved in AoSA. This approximation increases to around $98\%$ when $80$ phase shifters are employed. Regarding the soft max-min throughput optimization Algorithm 3, the approximations are around $70\%$ and $95\%$ when employing $32$ and $80$  phase shifters in nAoSA, respectively.} Clearly, there is only a modest  performance erosion {\color{black} when employing 80 phase shifters in  nAoSA, which is approximately half the number of phase shifters used in AoSA.} Furthermore, the minimum user throughputs obtained by the soft max-min throughput optimization Algorithm \ref{aalg3} are near to those obtained by the max-min throughput optimization Algorithm \ref{aalg1}.  Additionally, the users' minimum throughput achieved for $N_t = 2$ exhibits an improvement over the $N_t = 1$ scenario due to the increased spatial diversity gleaned at the UEs.

In Fig. \ref{fig:AoSA_nAoSA_P_sum_rate_vs_Lc}, we compare the sum-throughput performances of the soft max-min throughput optimization Algorithm \ref{aalg3} against the  sum-throughput maximization Algorithm \ref{aalg2}, evaluated at $P = 100$ mW. The sum-throughput achieved by  the soft max-min throughput maximization Algorithm \ref{aalg3} approaches that obtained by the  sum-throughput maximization Algorithm \ref{aalg2}.  Similar to the observations inferred  from Fig. \ref{fig:AoSA_nAoSA_P_min_rate_vs_Lc}, {\color{black} employing $32$ phase shifters in nAoSA achieves approximately $88\%$ of the sum-throughput obtained by AoSA,} when using $80$ phase shifters,   nAoSA  achieves approximately $95\%$ of the sum-throughput obtained by AoSA.

\begin{table*}[!t]
	\centering
	\caption{Total power consumption of nAoSA under different numbers of phase shifters $N_{PS}$ and AoSA with $N_{PS} = 144$ ($N_c = 8$, $P = 100$ mW)}
	\begin{tabular}{|l|cccc|}
		\hline
		& \multicolumn{1}{c|}{$N_{PS} = 32$} & \multicolumn{1}{c|}{$N_{PS} = 48$} & \multicolumn{1}{c|}{$N_{PS} = 64$} & $N_{PS} = 80$ \\ \hline
		{\sf nAoSA} & \multicolumn{1}{c|}{1684 mW} & \multicolumn{1}{c|}{2004 mW} & \multicolumn{1}{c|}{2324 mW} & 2644 mW \\ \hline
		{\sf AoSA} ($N_{PS} = 144$) & \multicolumn{4}{c|}{3924 mW} \\ \hline
	\end{tabular}
	\label{table:AoSA_nAoSA_Ptotal_vs_N_PS}
\end{table*}

\begin{table*}[!t]
	\centering
	\caption{Transmit power allocation by nAoSA under different numbers of phase shifters $N_{PS}$ and AoSA with $N_{PS} = 144$ ($N_c = 8$, $P_{total} = 3924$ mW)}
	\begin{tabular}{|l|cccc|}
		\hline
		& \multicolumn{1}{c|}{$N_{PS} = 32$} & \multicolumn{1}{c|}{$N_{PS} = 48$} & \multicolumn{1}{c|}{$N_{PS} = 64$} & $N_{PS} = 80$ \\ \hline
		{\sf nAoSA} & \multicolumn{1}{c|}{2340 mW} & \multicolumn{1}{c|}{2020 mW} & \multicolumn{1}{c|}{1700 mW} & 1380 mW \\ \hline
		{\sf AoSA} ($N_{PS} = 144$) & \multicolumn{4}{c|}{100 mW} \\ \hline
	\end{tabular}
	\label{table:AoSA_nAoSA_P_vs_N_PS}
\end{table*}

\subsection{Surpassing the AoSA performance by the nAoSA at equal total power}\label{sim_subsec3}
Next, we demonstrate the superiority of  nAoSA  in comparison to the conventional AoSA  under a fixed total power budget. This advantage originates from the controlled number of phase shifters
employed by our new structure for AP, allowing for a larger portion of power to be allocated to DP. To clarify, we define the total power as $P_{total} = P + N_c \times 118 + N_{PS} \times 20$, where $P$ represents the transmit power, $N_{PS}$ represents the number of phase shifters, and the power consumption per RF chain is set to $118$ mW, while the power consumption per phase shifter is set to $100$ mW \cite{LL16}. In the AoSA structure associated with $N_{PS} = 144$, $N_c=8$ and $P=100$ mW, following the parameter settings used in the aforementioned simulations, we establish a reference total power $P_{total}$ of $3924$ mW. By reducing $N_{PS}$ in the nAoSA AP, we assign a power budget exceeding $1000$ mW to DP, which in turn leads to enhanced users' throughputs.

We commence by evaluating the users'  minimum throughput performance{\color{black}, and compare it to the result of the max-min throughput optimization Algorithm \ref{aalg1} and to the soft max-min throughput optimization Algorithm \ref{aalg3}}  under the same total power budget, as depicted in Fig. \ref{fig:AoSA_nAoSA_Ptotal_min_rate_vs_Lc}. Given the similarity in performance between the nAoSA structure and the AoSA structure at the same transmit power, the former significantly outperforms the latter at the same total power.
When analyzing the sum-throughput {\color{black} achieved by the soft max-min throughput optimization Algorithm \ref{aalg3} and the sum-throughput maximization Algorithm \ref{aalg2}} under an equivalent total power, as depicted in Fig. \ref{fig:AoSA_nAoSA_nFC_Ptotal_sum_rate_vs_Lc}, an important observation can be made{\color{black}. Specifically, the} {\sf nAoSA-SMM} algorithm has the potential to outperform  the {\sf nAoSA-ST} algorithm.

{\color{black} Table \ref{table:AoSA_nAoSA_Ptotal_vs_N_PS} presents the total power $P_{total}$ necessary for simulating Fig. 3 and Fig. 4 along with $8$ RF chains and a transmit power $P$ of $100$ mW, comparing the nAoSA structure associated with varying number of phase shifters to the AoSA structure having an excessive number of $144$ phase shifters. Increasing the number of phase shifters in nAoSA allows for a closer approximation of AoSA in terms of both the minimum throughput and sum-throughput. However, this improvement comes at the expense of an increased total power budget, albeit still lower than that required by AoSA. Additionally, Table \ref{table:AoSA_nAoSA_P_vs_N_PS} presents the transmit power $P$ assigned to DP by nAoSA and AoSA in simulating Fig. 5 and Fig. 6 along with $8$ RF chains and a total power $P_{total}$ of $3924$ mW. The transmit power $P$ achieved by nAoSA significantly surpasses that achieved by AoSA, showing its potential to enhance the user throughput by reducing the circuit power at the AP and allocating additional power to DP.}

To clarify the observations gleaned from Fig. \ref{fig:AoSA_nAoSA_nFC_Ptotal_sum_rate_vs_Lc}, the distribution patterns of the individual user throughputs {\color{black} achieved by Algorithm \ref{aalg1}-\ref{aalg3},} for $N_t = 1$ and $N_t = 2$, using $80$ phase shifters and a total power of $3924$ mW are illustrated in Fig. \ref{fig:nAoSA_rate_distrib_Lc10_Ptotal}. It is worth noting that: $(i)$ When  sufficient transmit power is available, the soft max-min throughput optimization Algorithm \ref{aalg3}\footnote{As the transmit power budget $P$ for nAoSA DP exceeds $1000$ mW, given a total power of $3924$ mW, $\delta=1$ performs best}
manages to achieve a notable sum-throughput gain by realizing increased and balanced individual throughputs among the UEs; $(ii)$ Through conventional sum-throughput maximization, certain UEs may be assigned near-zero throughputs, rendering the sum-throughput algorithm unsuitable for multi-user services. Although maximizing the sum-throughput assigns higher throughputs to UEs with favorable channel conditions, the overall throughput loss caused by zero throughput allocations can occasionally outweigh the gains, which results in a diminished sum-throughput compared to that achieved by the soft max-min thoughput optimization Algorithm \ref{aalg3}. Therefore, the latter can be viewed as a beneficial near Pareto  solutions  simultaneously satisfying users' minimum throughput and sum-throughput targets.

\begin{figure}[!t]
	\centering
	\includegraphics[width=0.76\linewidth]{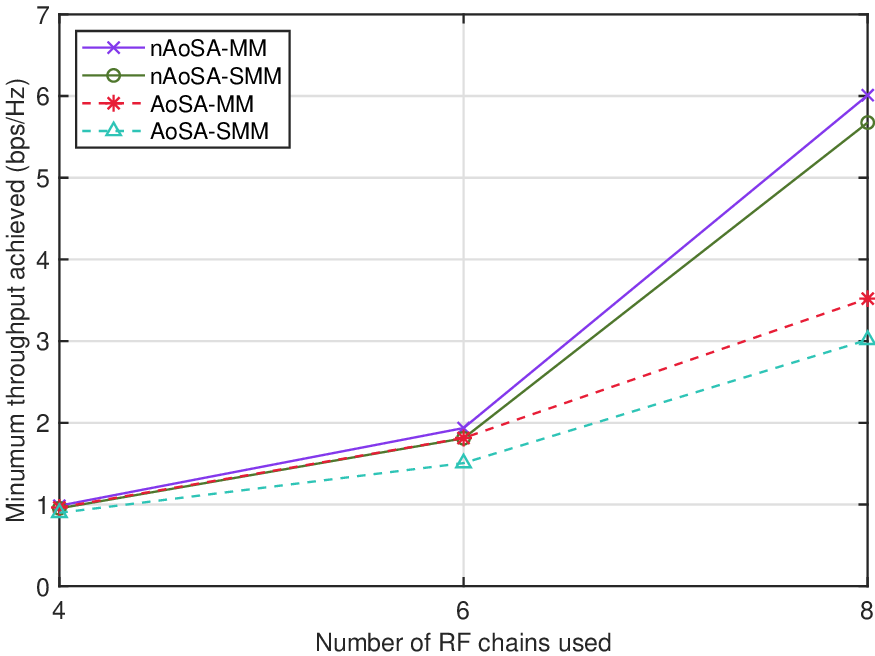}
	\caption{The minimum throughput achieved by nAoSA and AoSA vs. the number of RF chains used
		at equal total power $P_{total}$ for $N_t = 2$ (number of phase shifters used by AoSA and nAoSA is fixed at 144 and 72, respectively).}
	\label{fig:AoSA_nAoSA_Ptotal_min_rate_vs_Nc_Nt2}
\end{figure}

\begin{figure}[!t]
	\centering
	\includegraphics[width=0.76\linewidth]{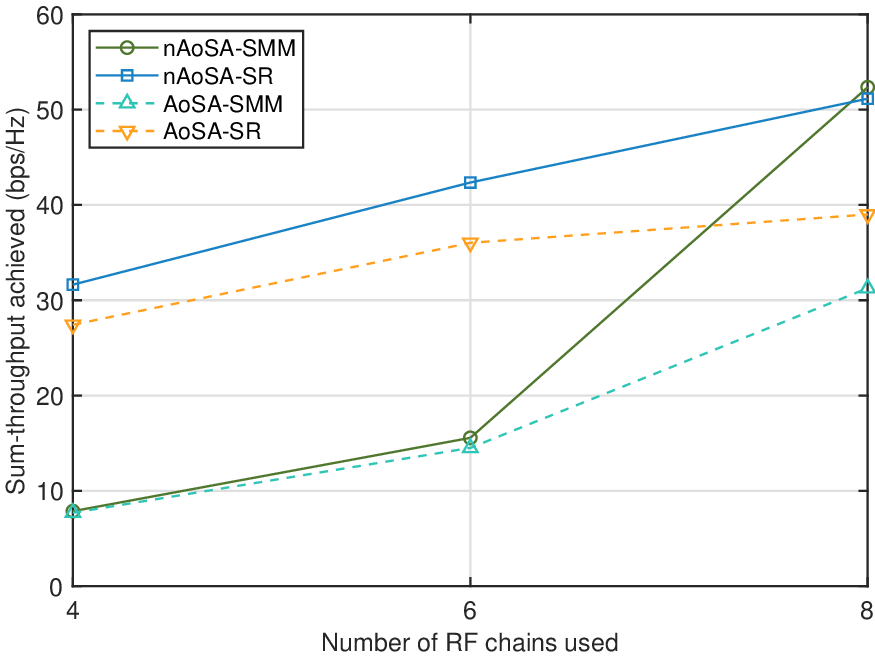}
	\caption{The sum-throughput achieved by nAoSA and AoSA vs. the number of RF chains used
		at equal total power $P_{total}$ for $N_t = 2$ (number of phase shifters used by AoSA and nAoSA is fixed at 144 and 72, respectively).}
	\label{fig:AoSA_nAoSA_Ptotal_sum_rate_vs_Nc_Nt2}
\end{figure}

{\color{black} Lastly, we evaluate the performance in terms of the minimum throughput and sum-throughput for different number of RF chains and $N_t = 2$. The AoSA structure uses $144$ phase shifters, while the the nAoSA structure uses $72$ phase shifters, allowing for the implementation of $4$, $6$ and $8$ RF chains. These assessments are conducted under a fixed total power of $P_{total} = 3924$ mW to highlight the trade-off between the number of RF chains and the transmit power $P$. Fig. \ref{fig:AoSA_nAoSA_Ptotal_min_rate_vs_Nc_Nt2} compares the minimum throughput achieved by the max-min-based Algorithm \ref{aalg1} and the soft max-min-based Algorithm \ref{aalg3}. With $8$ RF chains, the minimum throughput significantly outperforms that achieved with $4$ and $6$ RF chains, despite the latter having higher transmit power $P$. This demonstrates the benefits of employing more RF chains to enhance DP. Furthermore, Fig. \ref{fig:AoSA_nAoSA_Ptotal_sum_rate_vs_Nc_Nt2} compares the sum-throughput achieved by the soft max-min-based Algorithm \ref{aalg3} and the sum-throughput maximization Algorithm \ref{aalg2}. Similar to the observations gleaned from Fig. \ref{fig:AoSA_nAoSA_Ptotal_min_rate_vs_Nc_Nt2}, employing more RF chains leads to increased sum-throughput, even under reduced transmit power $P$.}

\section{Conclusions}
In response to the challenge of excessive power consumption associated with the state-of-the-art analog precoders (APs) used in large antenna-array-aided base stations for delivering multiple information streams to multi-antenna users, this paper has proposed a novel AP structure, which relies on a judiciously controlled number of low-resolution phase shifters. Based on the new AP, we have developed an optimization algorithm for designing hybrid precoders to maximize the users' minimum throughput for ensuring their quality-of-delivery. Furthermore, we conceived
 a new framework of optimizing  hybrid precoders by relying on sophisticated computational solutions. This framework achieves
a similar users' minimum throughput as to that obtained by directly maximizing the users' minimum throughput as well as a sum-throughput matching  that obtained by directly maximizing
the sum-throughput.
\section*{Appendix: Mathematical ingredient}
{\color{black}
Recall that \cite[p. 366]{Tuybook},
a function $\bar{f}$ is said to be a tight minorant (tight majorant, resp.) of a function $f$ over the domain $\mbox{dom}(f)$ at
 a point $\bar{z}\in\mbox{dom}(f)$, if it satisfies the conditions of global bounding
$f(z)\geq \bar{f}(z)$ $\forall\ z\in\mbox{dom}(f)$   ($f(z)\leq \bar{f}(z)$ $\forall\ z\in\mbox{dom}(f)$, resp.) and matching at $\bar{z}$: $f(\bar{z})=\bar{f}(\bar{z})$.

The following inequality  for all
$\bX$ and $\bar{X}$ of size $n\times m$ and $\bY\succ 0$ and $\bar{Y}\succ 0$ of size $n\times n$ has been established in \cite{TTN16,Tuaetal17}:
\begin{align}\label{fund5}
{\color{black}\ln\left|I_n+[\bX]^2\mathbf{Y}^{-1}\right|} \geq& {\color{black}\ln\left|I_n+[\bar{X}]^2\bar{Y}^{-1}\right|}
-\la [\bar{X}]^2\bar{Y}^{-1}\ra \nonumber\\
&+ 2\Re\{\la \bar{X}^H\bar{Y}^{-1}\bX\ra\}\notag \\
&- \big\la \bar{Y}^{-1}-(\bar{Y}+[\bar{X}]^2)^{-1},[\bX]^2+\mathbf{Y} \big\ra.
\end{align}
For
\begin{equation}\label{ap31a}
\Pi(\bX,\bY)\triangleq \sum_{k\in\clK}(I_{N_t}-\bX^H_k\bY^{-1}_k\bX_k),
\end{equation}
in the domain constrained by
\begin{eqnarray}\label{ap32}
\left\{ [\bX_k]^2\prec \bY_k, k=1,\dots, K\right\},
\end{eqnarray}
the following inequality holds for all $(\bX,\bY)$ and $(\bar{X},\bar{Y})$ \cite{Tuaetal23}
\begin{align}
\ln \Pi(\bX,\bY)\leq&\ln\Pi(\bar{X},\bar{Y})+\sum_{k\in\clK}\la \Pi^{-1}(\bar{X},\bar{Y})\bar{X}^H_k\bar{Y}_k^{-1}\bar{X}_k\ra\nonumber\\
&-2\sum_{k\in\clK}\Re\{\la \Pi^{-1}(\bar{X},\bar{Y}) \bar{X}^H_k\bar{Y}_k^{-1}\bX_k\ra\}\nonumber\\
&+\sum_{k\in\clK}\la \bar{Y}_k^{-1}\bar{X}_k \Pi^{-1}(\bar{X},\bar{Y})\bar{X}^H_k\bar{Y}_k^{-1}\bY_k\ra.\label{ap6}
\end{align}
Considering both sides of (\ref{fund5}) ((\ref{ap6}), resp.) as functions of the variables $(\bX,\bY)$, they
match at $(\bar{X},\bar{Y})$, i.e.
the function defined by the RHS  provides a tight minorant (tight majorant, resp.) of the log-determinant function  defined by the LHS at $(\bar{X},\bar{Y})$.
\bibliographystyle{IEEEtran}
\balance \bibliography{mmwave}

\begin{thebibliography}{10}
\providecommand{\url}[1]{#1}
\csname url@samestyle\endcsname
\providecommand{\newblock}{\relax}
\providecommand{\bibinfo}[2]{#2}
\providecommand{\BIBentrySTDinterwordspacing}{\spaceskip=0pt\relax}
\providecommand{\BIBentryALTinterwordstretchfactor}{4}
\providecommand{\BIBentryALTinterwordspacing}{\spaceskip=\fontdimen2\font plus
\BIBentryALTinterwordstretchfactor\fontdimen3\font minus \fontdimen4\font\relax}
\providecommand{\BIBforeignlanguage}[2]{{%
\expandafter\ifx\csname l@#1\endcsname\relax
\typeout{** WARNING: IEEEtran.bst: No hyphenation pattern has been}%
\typeout{** loaded for the language `#1'. Using the pattern for}%
\typeout{** the default language instead.}%
\else
\language=\csname l@#1\endcsname
\fi
#2}}
\providecommand{\BIBdecl}{\relax}
\BIBdecl

\bibitem{Rapetal19}
{T. S. Rappaport {\it et al.} }, ``Wireless communications and applications above 100 {GHz}: Opportunities and challenges for {6G} and beyond,'' \emph{IEEE Access}, vol.~7, pp. 78\,792--78\,757, 2019.

\bibitem{SN11}
H.-J. Song and T.~Nagatsuma, ``Present and future of terahertz communications,'' \emph{IEEE Trans. Terahertz Science Techn.}, vol.~1, no.~1, pp. 256--263, Jan. 2011.

\bibitem{Akyetal22}
I.~F. Akyildiz, C.~Han, Z.~Hu, S.~Nie, and J.~M. Jornet, ``Terahertz band communication: An old problem revisited and research directions for the next decade,'' \emph{IEEE Trans. Commun.}, vol.~70, no.~6, pp. 4250--4285, June 2022.

\bibitem{TD23}
J.~Tan and L.~Dai, ``{THz} precoding for {6G}: Challenges, solutions, and opportunities,'' \emph{IEEE Wirel. Commun.}, vol.~30, no.~4, pp. 132--138, Apr. 2023.

\bibitem{Gaoetal16}
X.~Gao, L.~Dai, S.~Han, C.-L. I, and R.~W. Heath, ``Energy-efficient hybrid analog and digital precoding for {MmWave} {MIMO} systems with large antenna arrays,'' \emph{IEEE J. Sel. Areas Commun.}, vol.~34, no.~4, pp. 998--1009, Apr. 2016.

\bibitem{Winetal13}
W.~Shin, O.~Inac, Y.-C. Ou, B.~Ku, and G.~M. Rebeiz, ``A {108-114 GHz 4x4} wafer-scale phased array transmitter with high-efficiency on-chip antennas,'' \emph{IEEE J. Solid-State Circuits}, vol.~48, no.~9, pp. 2041--2055, May 2013.

\bibitem{Ayetal14}
O.~{El Ayach}, S.~Rajagopal, S.~{Abu-Surra}, Z.~Pi, and R.~W. Heath, ``Spatially sparse precoding in millimeter wave {MIMO} systems,'' \emph{IEEE Tran. Wirel. Commun.}, vol.~13, no.~3, pp. 1499--1513, Mar. 2014.

\bibitem{Yuetal16}
X.~Yu, J.~C. Shen, J.~Zhang, and K.~B. Letaief, ``Alternating minimization algorithms for hybrid precoding in milimeter wave {MIMO} systems,'' \emph{IEEE J. Select. Topics Signal Process.}, vol.~10, no.~3, pp. 485--500, Apr. 2016.

\bibitem{LLW17}
C.~Lin, G.~Y. Li, and L.~Wang, ``Subarray-based coordinated beamforming training for {mmWave} and {sub-THz} communications,'' \emph{IEEE J. Sel. Areas Commun.}, vol.~35, no.~9, pp. 2115--2126, Sept. 2017.

\bibitem{YHY20}
L.~Yan, C.~Han, and J.~Yuan, ``A dynamic array-of-subarrays architecture and hybrid precoding algorithms for terahertz wireless communications,'' \emph{IEEE J. Select. Areas Commun.}, vol.~38, no.~9, pp. 2041--2056, Sep. 2020.

\bibitem{DTCP22}
L.~Dai, J.~Tan, Z.~Chen, and H.~V. Poor, ``Delay-phase precoding for wideband {THz} massive {MIMO},'' \emph{IEEE Trans. Wirel. Commun.}, vol.~21, no.~9, pp. 7271--7286, Sept. 2022.

\bibitem{SDN23}
R.~Su, L.~Dai, and D.~W.~K. Ng, ``Wideband precoding for {RIS}-aided {THz} communications,'' \emph{IEEE Trans. Commun.}, vol.~71, no.~6, pp. 3592--3604, June 2023.

\bibitem{Chenetal23}
Y.~Chen, J.~Tan, M.~Hao, R.~MacKenzie, and L.~Dai, ``Accurate beam training for {RIS}-assisted wideband terahertz communication,'' \emph{IEEE Trans. Commun. (early access)}, 2023.

\bibitem{SY16}
F.~Sohrabi and W.~Yu, ``Hybrid digital and analog beamforming design for large-scale antenna arrays,'' \emph{IEEE J. Selec. Topics Signal Process.}, vol.~10, no.~3, pp. 501--513, Mar. 2016.

\bibitem{KHY18}
L.~Kong, S.~Han, and C.~Yang, ``Hybrid precoding with rate and coverage constraints for wideband massive {MIMO} systems,'' \emph{IEEE Trans. Wirel. Commun.}, vol.~17, no.~7, pp. 4634--4647, Jul. 2018.

\bibitem{Shi-18-Jun-A}
Q.~{Shi} and M.~{Hong}, ``Spectral efficiency optimization for millimeter wave multiuser {MIMO} systems,'' \emph{IEEE J. Select. Topics Signal Process.}, vol.~12, no.~3, pp. 455--468, Jun. 2018.

\bibitem{Nasetal20TVT}
A.~A. Nasir, H.~D. Tuan, T.~Q. Duong, H.~V. Poor, and L.~Hanzo, ``Hybrid beamforming for multi-user millimeter-wave networks,'' \emph{IEEE Trans. Vehic. Techn.}, vol.~69, no.~3, pp. 2943--2956, Mar. 2020.

\bibitem{FMLS21}
C.~Fang, B.~Makki, J.~Li, and T.~Svensson, ``Hybrid precoding in cooperative millimeter wave networks,'' \emph{IEEE Trans. Wirel. Commun.}, vol.~20, no.~8, pp. 5373--5388, Aug. 2021.

\bibitem{SAA21}
H.~Sarieddeen, M.-S. Alouni, and T.~Y. Al-Naffouri, ``An overview of signal processing techniques for terahertz communications,'' \emph{Proc. IEEE}, vol. 109, no.~10, pp. 1628--1665, Oct. 2021.

\bibitem{Yuetal23tvt}
H.~Yu, H.~D. Tuan, E.~Dutkiewicz, H.~V. Poor, and L.~Hanzo, ``Low-resolution hybrid beamforming in millimeter-wave multi-user systems,'' \emph{IEEE Trans. Vehic. Techn.}, vol.~72, no.~7, pp. 8941--8955, Jul. 2023.

\bibitem{Yuetal23twc}
------, ``Regularized zero-forcing aided hybrid beamforming for millimeter-wave multiuser {MIMO} systems,'' \emph{IEEE Trans. Wirel. Commun.}, vol.~22, no.~5, pp. 3280--3295, May 2023.

\bibitem{AHRP13}
O.~E. Ayach, R.~W. Heath, S.~Rajagopal, and Z.~Pi, ``Multimode precoding in millimeter wave {MIMO} transmitters with multiple antenna sub-arrays,'' in \emph{Proc. IEEE Global Commun. Conf. (GLOBECOM)}, Dec. 2013, pp. 3476--3480.

\bibitem{KYW13}
J.~D. Krieger, C.-P. Yeang, and G.~W. Wornell, ``Dense delta-sigma phased arrays,'' \emph{IEEE Trans. Antenn. Propag.}, vol.~61, no.~4, pp. 1825--1837, Apr. 2013.

\bibitem{Betal06}
J.~F. Bonnans, J.~C. Gilbert, C.~Lemarechal, and C.~Sagastigabal, \emph{Numerical Optimization-Theoretical and Practical Aspects (second edition)}.\hskip 1em plus 0.5em minus 0.4em\relax Springer, 2006.

\bibitem{PTKD12}
A.~H. Phan, H.~D. Tuan, H.~H. Kha, and D.~T. Ngo, ``Nonsmooth optimization for efficient beamforming in cognitive radio multicast transmission,'' \emph{IEEE Trans. Signal Process.}, vol.~60, pp. 2941--2951, Jun. 2012.

\bibitem{CTN14}
E.~Che, H.~D. Tuan, and H.~H. Nguyen, ``Joint optimization of cooperative beamforming and relay assignment in multi-user wireless relay networks,'' \emph{IEEE Trans. Wirel. Commun.}, vol.~13, no.~10, pp. 5481--5495, Oct. 2014.

\bibitem{Tametal16}
H.~H.~M. Tam, H.~D. Tuan, D.~T. Ngo, T.~Q. Duong, and H.~V. Poor, ``Joint load balancing and interference management for small-cell heterogeneous networks with limited backhaul capacity,'' \emph{IEEE Trans. Wirel. Commun.}, vol.~16, no.~2, pp. 872--884, Feb. 2017.

\bibitem{Yeetal20}
Y.~Shi, H.~D. Tuan, T.~Q. Duong, H.~V. Poor, and A.~V. Savkin, ``{PMU} placement optimization for efficient state estimation in smart grid,'' \emph{IEEE J. Sel. Areas Commun.}, vol.~38, no.~1, pp. 71--83, Jan. 2020.

\bibitem{Yuetal20jsac}
H.~Yu, H.~D. Tuan, A.~A. Nasir, T.~Q. Duong, and H.~V. Poor, ``Joint design of reconfigurable intelligent surfaces and transmit beamforming under proper and improper {G}aussian signaling,'' \emph{IEEE J. Select. Areas Commun.}, vol.~38, no.~11, pp. 2589--2603, Nov. 2020.

\bibitem{Tuaetal23}
H.~D. Tuan, A.~A. Nasir, E.~Dutkiewicz, H.~V. Poor, and L.~Hanzo, ``{RIS}-aided multiple-input multiple-output broadcast channel capacity,'' \emph{IEEE Trans. Commun. (early access)}.

\bibitem{EMMZ20}
A.~Epasto, M.~Mahdian, V.~Mirrokni, and M.~Zampetakis, ``Optimal approximation-smoothness tradeoffs for soft-max functions,'' in \emph{Adv. Neural Inf. Process. Syst.}, 2020, pp. 1--10.

\bibitem{Akd14}
M.~R. Akdeniz, Y.~Liu, M.~K. Samimi, S.~Sun, S.~Rangan, T.~S. Rappaport, and E.~Erkip, ``Millimeter wave channel modeling and cellular capacity evaluation,'' \emph{IEEE J. Select. Areas Commun.}, vol.~32, no.~6, pp. 1164--1179, 2014.

\bibitem{Linetal21}
{Z. Lin et al.}, ``Joint estimation of multipath angles and delays for millimeter-wave cylindrical arrays with hybrid front-ends,'' \emph{IEEE Trans. Wirel. Commun.}, vol.~20, no.~7, pp. 4631--4645, Jul. 2021.

\bibitem{LL16}
C.~Lin and G.~Y. Li, ``Energy-efficient design of indoor {mmWave} and {Sub-THz} systems with antenna arrays,'' \emph{IEEE Trans. Wirel. Commun.}, vol.~15, no.~7, pp. 4660--4672, Jul. 2016.

\bibitem{Tuybook}
H.~Tuy, \emph{Convex Analysis and Global Optimization (second edition)}.\hskip 1em plus 0.5em minus 0.4em\relax Springer International, 2017.

\bibitem{TTN16}
H.~H.~M. Tam, H.~D. Tuan, and D.~T. Ngo, ``Successive convex quadratic programming for quality-of-service management in full-duplex {MU-MIMO} multicell networks,'' \emph{IEEE Trans. Commun.}, vol.~64, no.~6, pp. 2340--2353, Jun. 2016.

\bibitem{Tuaetal17}
H.~D. Tuan, H.~H.~M. Tam, H.~H. Nguyen, T.~Q. Duong, and H.~V. Poor, ``Superposition signaling in broadcast interference networks,'' \emph{IEEE Trans. Commun.}, vol.~65, no.~11, pp. 4646--4656, Nov. 2017.

\end{thebibliography}

\end{document}